\documentclass[pre,twocolumn,aps,10pt,superscriptaddress,notitlepage,nofootinbib]{revtex4}  
\usepackage{epsfig,amsmath,amssymb,amsfonts,graphicx,color,calc,epstopdf,stmaryrd}

\let\a=\alpha \let\b=\beta \let\g=\gamma \let\d=\delta
\let\e=\varepsilon \let\z=\zeta \let\h=\eta 
\let\l=\lambda \let\m=\mu  \let\x=\xi 
\let\s=\sigma \let\t=\tau \let\f=\varphi \let\c=\chi
 \let\y=\upsilon  \let\G=\Gamma
\let\D=\Delta \let\Th=\Theta\let\L=\Lambda  
\let\Si=\Sigma   
\let\ee=\epsilon \let\r=\rho \let\th=\theta \let\io=\infty

\def\FF{{\cal F}} 
\def\NN{{\cal N}} 
\def\LL{{\cal L}}  
\def\DD{{\cal D}}

\def\to{\rightarrow} \def\la{\left\langle} \def\ra{\right\rangle}

\def\ol{\overline}

\def\wh{\widehat}

\def\de{\mathrm d}

\newcommand{\beq}{\begin{equation}} 
\newcommand{\eeq}{\end{equation}}
\newcommand{\ba}{\begin{eqnarray}}
\newcommand{\ea}{\end{eqnarray}}

\def\SIcite{the Appendix}

\newcommand{\afunc}[1]{\operatorname{\mathsf{#1}}}
\def\DE{\afunc{D}}
\def\de{\mathrm d}

\def\Dg{\Delta^g}
\def\De{\Delta}

\begin{document}

\title{
Following the evolution of glassy states under external perturbations: \\ compression and shear-strain
}

 \author{Corrado Rainone}
 \affiliation{LPT,
 Ecole Normale Sup\'erieure, CNRS UMR 8549, 24 Rue Lhomond, 75005 France}
\affiliation{Dipartimento di Fisica,
Sapienza Universit\`a di Roma,
P.le A. Moro 2, I-00185 Roma, Italy}

\author{Pierfrancesco Urbani}
\affiliation{IPhT, CEA/DSM-CNRS/URA 2306, CEA Saclay, F-91191 Gif-sur-Yvette Cedex, France}

\author{Hajime Yoshino}
\affiliation{Cybermedia Center, Osaka University, Toyonaka, Osaka 560-0043, Japan}
\affiliation{Graduate School of Science, Osaka University, Toyonaka, Osaka 560-0043, Japan}

 \author{Francesco Zamponi}
 \affiliation{LPT,
 Ecole Normale Sup\'erieure, CNRS UMR 8549, 24 Rue Lhomond, 75005 France}

\begin{abstract}
We consider the adiabatic evolution of glassy states under external perturbations.
Although the formalism we use is very general, we focus here on infinite-dimensional hard spheres where
an exact analysis is possible.
We consider perturbations of the boundary, i.e. compression or (volume preserving)
shear-strain, and we compute the response of glassy states to such perturbations: pressure
and shear-stress. We find that both quantities overshoot before the glass
state becomes unstable at a spinodal point where it melts into a liquid (or yields). 
We also estimate the yield
stress of the glass. Finally, we study the stability of the glass basins towards breaking into
sub-basins, corresponding to a Gardner transition. We find that close to the dynamical transition,
glasses undergo a Gardner transition after an infinitesimal perturbation.
\end{abstract}

\maketitle

\paragraph*{Introduction --}
Glasses are long lived metastable states of matter, in which particles are confined around an amorphous structure~\cite{Ca09,Dy06}.
For a given sample of a material, the glass state is not unique: 
depending on the preparation protocol, the material can be trapped in different glasses, 
each displaying different thermodynamic properties.
For example, 
the specific volume of a glass prepared by cooling a liquid
depends strongly on the cooling rate~\cite{Ca09,Dy06}.
Other procedures, such as vapor deposition, produce very stable glasses, with higher
density than those obtained by simple cooling~\cite{Sw07,SEP13}. When heated up,
glasses show hysteresis: their energy (specific volume) remains below the
liquid one, until a ``spinodal'' point is reached, at which they melt into the liquid (see
e.g.~\cite[Fig.1]{Dy06} and \cite[Fig.2]{SEP13}). 

The behavior of glasses under shear-strain also shows similarly complex phenomena.
Suppose to prepare a glass by cooling a liquid at a given rate until some low temperature $T$ is reached.
After cooling, a strain $\g$ is applied and the stress $\s$ is recorded. At small $\g$, an elastic (linear) regime
where $\s \sim \mu \g$ is found. At larger $\g$, the stress reaches a maximum and then decreases until
an instability is reached, where the glass yields and starts to flow (see e.g.~\cite[Fig.3c]{RTV11} and~\cite[Fig.2]{Kou12}). 
The amplitude of the shear modulus $\mu$ and of the stress overshoot increase when the cooling rate
is decreased, and more stable glasses are reached.

Computing these observables theoretically is a difficult challenge, because
glassy states are always prepared through non-equilibrium dynamical protocols.
First-principle dynamical theories such as Mode-Coupling Theory (MCT)~\cite{Go09} are successful
in describing properties of supercooled liquids close to the glass state 
(including the stress overshoot~\cite{Br09}), 
but they fail to describe glasses at low temperatures and high pressures~\cite{IB13}.
The dynamical facilitation picture can successfully describe calorimetric properties of glasses~\cite{KGC13}, but
for the moment it does not allow one to perform 
first-principles calculations starting from the microscopic interaction potential.
To bypass the difficulty of describing all the dynamical details of glass formation, one can exploit a standard 
idea in statistical mechanics, namely that metastable states are described by a {\it restricted} equilibrium
thermodynamics for times much shorter than their lifetimes~\cite{PL71,La74}.
Within schematic models of 
glasses, this construction was proposed by several authors~\cite{KW87,KT89,Mo95,FP95}
and was formalised through the Franz-Parisi free energy~\cite{FP95} and the
``state following'' formalism~\cite{BFP97,KZ10b,KZ13}. 

In this paper we apply the state following construction~\cite{FP95,BFP97,KZ10b,KZ13} to a realistic model 
of glass former, made by identical particles interacting in the continuum. 
For simplicity, we choose here hard spheres in spatial dimension $d\to\io$, 
where the method is exact because metastable
states have infinite lifetime~\cite{KW87,PZ10,nature}.
We show that all the properties of glasses mentioned above are predicted by this framework,
including the cooling rate dependence of the specific volume (or the pressure)~\cite{Dy06,Ca09}, 
the hysteresis observed upon heating glasses~\cite{Dy06,Sw07,SEP13}, the behavior of the shear modulus
and the stress overshoot~\cite{RTV11,Kou12}.
Following~\cite{MP09,PZ10},
our method can be generalized (under standard liquid theory approximations) 
to experimentally relevant systems in $d=2,3$ with different interaction potentials, 
to obtain precise quantitative predictions, as we discuss in the conclusions.

\paragraph*{Constrained thermodynamics --} 
The ``state following'' formalism is designed to describe glass formation during slow cooling of a liquid~\cite{KZ13}.
Approaching the glass transition, the equilibrium dynamics of the liquid
happens on two well separated time scales~\cite{Dy06,Ca09}.
On a $T$-independent fast scale $\t_{\rm vib}$ particles
essentially vibrate in the cages formed by their neighbors.
On the slow $\a$-relaxation scale $\t_\a(T)$, that increases fast approaching the glass transition, 
cooperative processes change the structure 
of the material. 
When $\t_\a(T) \gg \t_{\rm vib}$, the system vibrates for a long time around a locally stable
configuration of the particles (a glass), and then on a time scale $\t_\a(T)$ transforms in another equivalent
glass. Hence, $\t_\a(T)$ is the lifetime of metastable glasses.
The liquid reaches equilibrium if enough different glass states are visited, hence
the experimental time scale (e.g. the cooling rate) should be $\t_{\rm exp} \gg \t_\a(T)$.
For given $\t_{\rm exp}$, the glass transition temperature $T_g$
 is therefore defined by $\t_{\rm exp}= \t_\a(T_g)$~\cite{Dy06,Ca09}. 
 For $T<T_g$ the system is confined into a given glass with lifetime $\t_\a(T) \gg \t_{\rm exp}$,
 which can thus be considered an infinitely-long lived metastable state. 
 Although the system is strictly speaking out of equilibrium in this regime, the slow relaxation is effectively 
 frozen and the material is confined in a thermodynamic equilibrium state {\it restricted} to a given glass.
In fact, if cooling stops at some $T < T_g$,
thermodynamic quantities quickly reach time-independent values, that satisfy equilibrium thermodynamic relations.
Still, the ``thermodynamic'' state depends on preparation history, and most crucially on the 
temperature $T_g$ at which the liquid fell out of equilibrium.
Note that aging effects can be neglected here because they happen, for $T < T_g$,
on time scales $\t_{\rm aging} \gg \t_\a(T_g) \sim \t_{\rm exp}$.

This observation suggests how 
to describe the thermodynamic properties of glasses
prepared by slow cooling~\cite{FP95,BFP97,KZ10b,KZ13}.
Consider $N$ interacting classical particles, described by coordinates 
$X = \{ x_i \}_{i=1,\cdots,N}$ and potential energy $V(X)$. 
During a cooling process with time scale $\t_{\rm exp}$,
the system remains equilibrated provided $T \geq T_g$. 
Define $R = \{ r_i \}$ the last configuration visited by the material before falling out of equilibrium;
its probability distribution is the equilibrium one at $T_g$, $P(R) = \exp[-V(R)/T_g]/Z(T_g)$ (here $k_B=1$).
For $T < T_g$, 
the lifetime of glasses becomes effectively infinite~\footnote{
A short transient when $\t_\a(T) \sim T_g$ exist, where the system is neither at equilibrium
nor confined in a glass. However, because $\t_\a(T)$ increases quickly around $T_g$, for slow coolings
this temperature regime is extremely small and negligible.}:
the material visits configurations $X$ confined in the glass selected by $R$. 
This constraint is implemented~\cite{FP95,BFP97} by imposing that the mean
square displacement between $X$ and $R$, $\D(X,R) = (d/N) \sum_{i=1}^N (x_i - r_i)^2$, be smaller than
a prescribed value $\D^{\rm r}$. 
The evolution of this glass is followed by changing
its temperature $T$ or applying some
perturbation $\g$ that changes the potential to $V_\g$.
The free energy of the glass selected by $R$ is therefore
\beq\nonumber
F_g[T,\g;R] = - T \log \int dX e^{-V_\g[X]/T} \th[ \D^{\rm r} - \D(X,R) ] \ .
\eeq
$\th(x)$ is the Heaviside function.
Computing $F_g[T,\g;R]$ is a formidably difficult task, because the constraint $\D(X,R) \leq \D^{\rm r}$ 
explicitly breaks translational
invariance and prevents one from using standard statistical mechanics methods.
One can simplify the problem
by computing the average free energy of all glasses that are sampled by liquid configurations at $T_g$,
under the assumption that these glasses have similar thermodynamic properties. We obtain
\beq\nonumber
F_g[T,\g] = \overline{F_g[T,\g;R]} = \int dR \frac{e^{-V(R)/T_g}}{Z(T_g)} F_g[T,\g;R] \ .
\eeq
This average can be computed using the replica trick~\cite{FP95}, and here we use the simplest
replica symmetric (RS) scheme~\cite{FP95,BFP97,KZ10b}.
The parameter $\D^{\rm r}$ is determined by minimizing the free energy, see~\SIcite.

\begin{figure}[t]
\includegraphics[width=\columnwidth]{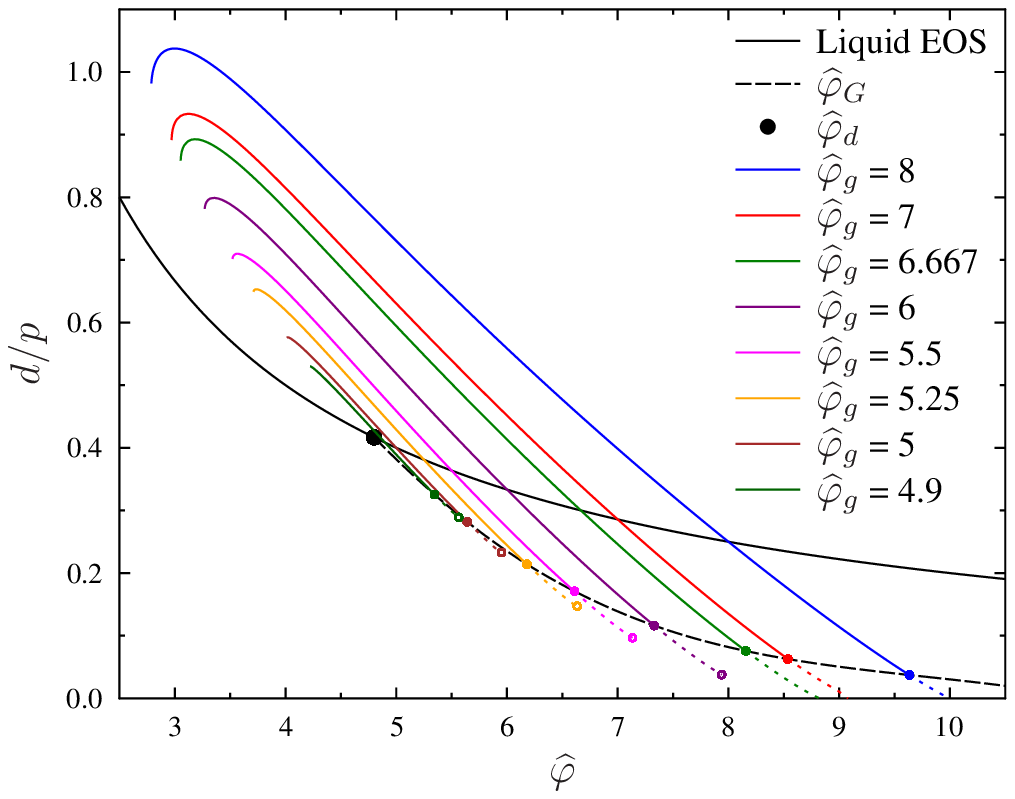}
\caption{Following glasses in (de)compression. Inverse 
reduced pressure $d/p$ is plotted
versus packing fraction $\wh\f = 2^d \f/d$. 
Both quantities are scaled to have a finite limit for $d\to\io$. 
The liquid EOS is $d/p = 2/\wh\f$. The dynamical transition
$\wh\f_{\rm d}$ is marked by a black dot. For $\wh\f_g > \wh\f_{\rm d}$, the liquid is a collection of glasses.
The glassy EOS are reported as full colored lines, that intersect the liquid EOS at $\wh\f_g$. Upon compression, 
a glass prepared at $\wh\f_g$ undergoes a Gardner transition at $\wh\f_G(\wh\f_g)$ (full symbols 
and long-dashed black line). Beyond $\wh\f_G$ our computation is not correct: glass EOS are reported 
as dashed lines. For low $\wh\f_g$ they end at an unphysical spinodal point (open symbol). 
Upon decompression, the glass pressure falls below the liquid one, until it reaches a minimum,
and then grows again until a physical spinodal point at which the glass melts into the liquid.
}
\label{fig:compression}
\end{figure}

\begin{figure}[t]
\includegraphics[width=.45\textwidth]{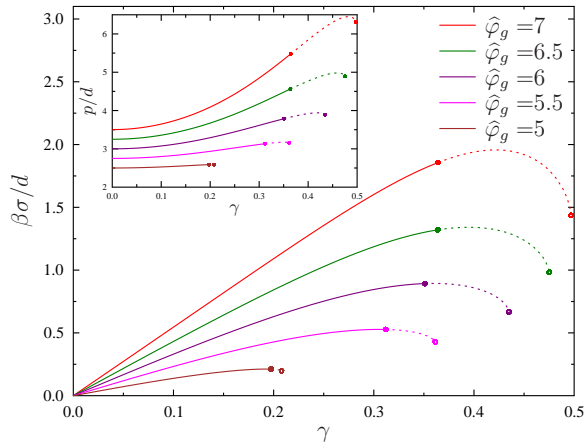}
\caption{Following glassy states prepared at $\wh\f_g$ upon applying a shear-strain $\g$.
Shear-stress $\s$ {\it (main panel)} and reduced pressure $p$ 
{\it (inset)} as a function of strain for different $\wh\f_g$.
Same styles as Fig.~\ref{fig:compression}. Upon increasing shear-strain,
the states undergo a Gardner transition at $\g_G(\wh\f_g)$. For $\g > \g_G$ our RS computation is
unstable but it predicts a stress overshoot followed by a spinodal point.
}
\label{fig:shear}
\end{figure}

This computation was done for spin glasses in~\cite{FP95,BFP97,KZ10b,KZ13,FPR14} and
describes perfectly the properties of glasses obtained by slow cooling~\cite{KZ13}. 
Here we consider a realistic glass-former:
a hard sphere system for $d\to\io$.
Technically, the computation uses the methods of~\cite{nature}
in the more complicated state following setting.
Because the details are not particularly instructive, we report them
in the Appendix, where we also discuss
the conceptual differences
with respect to previous works~\cite{PZ10,nature}.

\paragraph*{Results: compression --}
As a first application of the method, we consider preparing glasses by slow compression,
which is equivalent, for hard spheres, to slow cooling~\cite{PZ10}. Note that for hard spheres
temperature can be eliminated by appropriately rescaling physical quantities.
The system is prepared at low
density $\r$, particle volume $V_s$ is slowly increased (equivalently, container volume is decreased), and pressure $P$ is
monitored. In Fig.~\ref{fig:compression} we plot the reduced pressure $p = \b P/\r$, with $\b=1/T$, versus the 
packing fraction $\f = \r V_s$. At equilibrium, the system follows the liquid equation of state (EOS). Above the so-called
{\it dynamical transition} (or MCT transition) density
$\f_{\rm d}$, glasses appear, and equilibrium liquid configurations at $\f_g > \f_{\rm d}$ select a glass.
In Fig.~\ref{fig:compression} we report the EOS of several glasses corresponding to different choices of $\f_g$.
The slope of the glass EOS at $\f_g$ is different from that of the liquid EOS, indicating that when the system
falls out of equilibrium at $\f_g$, the compressibility has a jump, as observed experimentally~\cite{PZ10,CIPZ11}.
Following glasses in compression, pressure increases faster than in the liquid (compressibility is smaller)
and diverges at a finite {\it jamming} density $\f_j(\f_g)$~\cite{PZ10}. 
However, before jamming is reached, the glass undergoes a {\it Gardner transition}~\cite{Ga85,nature},
at which individual glass basins split in a fractal structure of subbasins. Because this transition was discussed 
before~\cite{Ga85,BFP97,KZ10b,nature}, we do not insist on its characterization, but note that we can compute precisely
the Gardner transition point $\f_G(\f_g)$ for all $\f_g$ (see~\SIcite~for details). 
Interestingly, as observed in~\cite{BFP97,FPR14}, the Gardner transition line ends at $\f_{\rm d}$, i.e. $\f_G(\f_g=\f_{\rm d}) = \f_{\rm d}$.
This implies that the first glasses appearing at $\f_{\rm d}$ are marginally stable towards breaking into subbasins, while glasses
appearing at $\f_g > \f_{\rm d}$ remain stable for a finite interval of pressures before breaking into subbasins. 
Yet, all glasses undergo the Gardner transition at finite pressure before jamming occurs~\cite{nature}.

\begin{figure}[t]
\includegraphics[width=.45\textwidth]{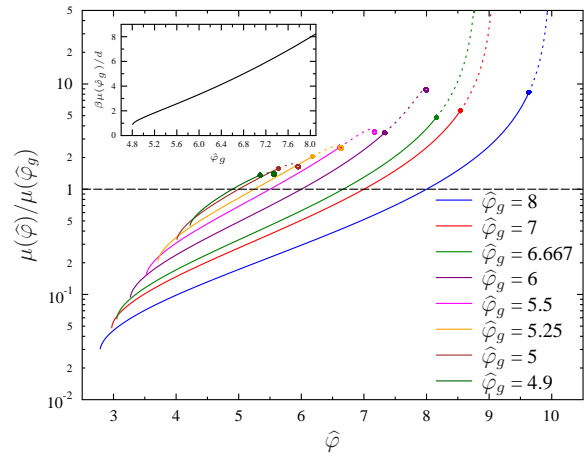}
\caption{
Shear modulus versus density for different glasses. Same styles as Fig.~\ref{fig:compression}.
In the inset we report $\mu(\wh\f_g)$ versus $\wh\f_g$. 
Note that the dilatancy $R/\rho=(1/2)\wh\f\partial \mu/\partial \wh\f$ diverges both at jamming and at
the low density spinodal point where the glass melts (see~\SIcite).
}
\label{fig:muR}
\end{figure}

For a glass selected at $\f_g$, when the density is higher than $\f_G$, the RS calculation we perform
is incorrect. One should perform a full replica symmetry breaking (fRSB) computation~\cite{BFP97,KZ10b,nature}. We leave this for future
work, but we observe that for large enough $\f_g$ the Gardner transition happens at very high pressure and in that case the RS calculation
should be a good approximation to the glass EOS at all pressures. For small $\f_g$ instead, the RS calculation gives a wrong prediction,
namely the existence of an unphysical spinodal point at which the glass disappears. We expect, based on the analogy 
with the results of~\cite{KZ10b}, that a fRSB calculation will fix this problem.

A given glass prepared at $\f_g$ can be also followed in {\it decompression}, by decompressing
at a relatively fast rate $\t_{\rm dec}$ such that $\t_{\rm vib} \ll \t_{\rm dec} \ll \t_{\rm exp}$.
In this case we observe hysteresis (Fig.~\ref{fig:compression}), consistently with experimental 
results~\cite{Dy06,Sw07,SEP13}. In fact, the glass pressure becomes {\it lower} than the liquid
one, until upon decreasing density a spinodal point is reached, at which the glass becomes 
unstable and melts into the liquid~\cite{MPR14}. Note that pressure ``undershoots'' (it has a local minimum, see Fig.~\ref{fig:compression}) 
before the spinodal is
reached~\cite{MPR14}.

\paragraph*{Results: shear --}
We investigate now the response of glasses to a shear-strain perturbation.
We consider a system compressed in equilibrium up to a density $\wh\f_g$,
where it remains stuck into a glass. Now, instead of compressing the system, 
we apply a shear-strain $\g$.
In Fig.~\ref{fig:shear} we report the behavior of shear-stress $\s$ and pressure $p$ versus $\g$.
At small $\g$ we observe a {\it linear response} elastic regime where 
$\s$ increases linearly with $\g$, $\s \sim \mu \g$
and pressure increases quadratically above the equilibrium liquid value, $p(\g) \sim p(\g=0) + (\b R/\r) \g^2$. 
Both the shear modulus $\mu$
and the dilatancy $R>0$ increase with $\wh\f_g$,
indicating that glasses prepared by slower annealing are more rigid. 

Upon further increasing $\g$, glasses enter a non-linear regime, and undergo a Gardner
transition at $\g_G(\f_g)$ (Fig.~\ref{fig:shear}). 
Like in compression, we find $\g_G(\f_{\rm d})=0$, and
$\g_G$ increases rapidly with $\f_g$. For $\g > \g_G(\f_{\rm d})$, the glass breaks into subbasins
and a fRSB calculation is needed. Note that the
RS computation predicts a stress overshoot, followed by a spinodal point where
the glass basin disappears. We expect that the fRSB computation gives similar results. The spinodal
point corresponds to the point where the glass yields and starts to flow. 
The values of yield strain $\g_{\rm Y}$ and of yield stress $\s_{\rm Y}$ are also found to increase with $\f_g$.
These results are qualitatively 
consistent with the experimental and numerical observations of~\cite{RTV11,Kou12}.

\paragraph*{Results: compression followed by shear --} 
One could also consider the case where {\it (i)} a liquid is slowly compressed up to $\f_g$ where it forms a glass, 
{\it (ii)} the glass is compressed up to a certain pressure $p$ (Fig.~\ref{fig:compression}) and then {\it (iii)} a shear-strain 
$\g$ is applied. The response to shear-strain of these glasses compressed out of equilibrium is qualitatively
similar to the one reported in Fig.~\ref{fig:shear}, and we do not report the corresponding curves. Instead,
we report in Fig.~\ref{fig:muR} the behavior of shear modulus $\mu$ as a function of density $\f$ for different
glasses prepared at different $\f_g$. For each glass, we find that under compression $\mu$
increases with density, and diverges at the jamming point where $p\to \io$. 
Note that, as discussed above and in~\cite{nature}, describing
the behavior around the jamming density requires a fRSB computation, that we did not perform here. 

A useful thermodynamic identity gives the dilatancy $R/\rho=(1/2)\f\partial \mu/\partial \f$~\cite{Ti14} (see~\SIcite).
This implies that the singular behavior of the shear modulus around jamming,  
which itself is well captured by a fRSB computation~\cite{YZ14},
should be directly reflected to the dilatancy, as pointed out in~\cite{Ti14}.
Further work is needed to understand experimental and numerical results~\cite{RDB13,CSD14,OH14}.

\paragraph*{Conclusions --}
We have applied the state following procedure, developed in the context of spin glasses~\cite{FP95,BFP97,KZ10b,KZ13},
to a microscopic model of glass former, namely hard spheres. We considered for simplicity the limit $d\to\io$, where
the method we used is exact, but the calculations can be generalized to obtain {\it approximated} 
quantitative predictions in finite $d$. According to~\cite{PZ10,BJZ11}, the simplest 
approximation is to use the results reported in this paper, replacing $\wh\f = 2^d \f / (d \, y^{\rm liq}_{\rm HS}(\f))$, 
$y^{\rm liq}_{\rm HS}(\f)$ being the contact value of the pair correlation function in the liquid phase, which can be obtained
from a generalized Carnahan-Starling liquid EOS~\cite{CIPZ11}. This approximation is expected to be
good at large $\f_g$, but gives poor results for $\f_g \sim \f_{\rm d}$. Systematic improvements over this approximation can
be obtained following the ideas of~\cite{PZ10}. It is clear, anyway, that the qualitative shape of the curves we obtained in $d\to\io$
will not change in finite $d$, which is also supported by the numerical simulations of~\cite{CIPZ11}.

We did not attempt here a more precise quantitative comparison with experimental and numerical data, which we leave for future work, but we
showed that the state following method is able to give predictions for many physical observables of experimental interest, and
reproduces a quite large number of observations. These include: {\it (i)} the pressure as a function of density for different glasses (Fig.~\ref{fig:compression}), which
displays a jump in compressibility at $\f_g$~\cite{PZ10,CIPZ11}; {\it (ii)} the presence of hysteresis and of a spinodal point in decompression
in the pressure-density curves (Fig.~\ref{fig:compression}), where we show that
more stable glasses (those with higher $\f_g$) display a larger hysteresis, consistently with the experimental observation of~\cite{Dy06,Sw07,SEP13};
the behavior of pressure and shear-stress under a shear-strain perturbation (Fig.~\ref{fig:shear}), where we show that {\it (iii)}
the shear modulus and the dilatancy increase for more stable glasses (higher $\f_g$), and {\it (iv)} that the shear-stress overshoots before a spinodal (yielding)
point is reached where the glass yields and starts to flow (Fig.~\ref{fig:shear})~\cite{RTV11,Kou12}. 
Note however that the spinodal (yield) point falls beyond the Gardner transition and therefore
its estimate, 
reported in Fig.~\ref{fig:shear}, is only approximate, a correct computation requires fRSB~\cite{nature}.
Furthermore, {\it (v)} we predict that glasses undergo a Gardner transition both in compression (Fig.~\ref{fig:compression}) and in shear (Fig.~\ref{fig:shear}), and
we locate the Gardner transition point (see~\SIcite).
Finally, we {\it (vi)} compute the dilatancy and the shear modulus everywhere in the glass phase (Fig.~\ref{fig:muR} and~\SIcite)
and their behavior close to the jamming transition (see~\SIcite).

This approach thus provides a coherent picture of the phase diagram of glasses in different regimes, under compression and under shear-strain, at moderate 
densities close to the dynamical glass transition and at high densities (pressures) close to jamming.
Future work should be directed towards performing 
systematic comparisons between theory and experiment,
and improving the theory, first by performing the fRSB computation, and second
by improving the approximation in finite dimensions.

\vskip10pt

\paragraph*{Acknowledgments --}
We thank G.~Biroli, 
P.~Charbonneau, O.~Dauchot, S.~Franz, Y.~Jin, F.~Krzakala,
J.~Kurchan, M.~Mariani, G.~Parisi, F.~Ricci-Tersenghi and L.~Zdeborov\'a, 
for many useful discussions.
This work was supported by KAKENHI (No. 25103005  ``Fluctuation \& Structure'' and No. 50335337)
from MEXT, Japan, by JPS Core-to-Core Program ``Non-equilibrium dynamics of soft matter and informations'',
and by the European Research Council through ERC grant agreement No. 247328 and NPRGGLASS.

\clearpage

\appendix

\begin{widetext}

\tableofcontents

\section{The Franz-Parisi free energy}

The Franz-Parisi potential allows one to compute the properties of an individual glassy state.
We consider a set of $m$ coupled ``reference'' replicas $R^1, \cdots, R^m$, where $R = \{ r_i \}$ is a configuration of the system.
These $m$ replicas interact with a potential energy $V(R) = \sum_{i<j} v(r_i - r_j)$, at temperature $T_g$ (with $k_B=1$), and
are used to select a glassy state, following~\cite{Mo95}.
At equilibrium we shall consider $m=1$, while out of equilibrium states can be selected using $m \neq 1$. In this paper we write the formulae
for general $m$, but we will only report results for $m=1$. 
Furthermore, we consider an additional ``constrained'' replica $X = \{ x_i \}$, which is coupled to one of the $m$ replicas and is used to probe
the glassy state selected by the reference replicas. This constrained replica has potential energy $V_{\g}(X) = \sum_{i<j} v_\g(x_i-x_j)$, where the suffix $\g$ is there to indicate
the possible presence of perturbations (e.g. a shear strain), and temperature $T$.
We define the mean square displacement as 
\beq
\D(X,R) = \frac{d}{N} \sum_{i=1}^N (x_i - r_i)^2 \ .
\eeq
The factor of $d$ is added to ensure that $\D(X,R)$ has a finite limit for $d\to\io$~\cite{PZ10,KPZ12,KPUZ13,CKPUZ13}.

In the following we restrict the discussion to a mean-field setting in which metastable states have infinite life-time and
phase coexistence is absent (see~\cite{MP00} for a discussion of phase coexistence in this context).
This is exactly realized in the limit $d\to\io$~\cite{PZ10,KPZ12,KPUZ13,CKPUZ13} and can be taken as an approximation for finite $d$.
We want to compute the free energy of the constrained replica, and average over
the reference replicas:
\beq
\begin{split}
Z_g[R,\D^r] &= \int dX  e^{-\b V_\g(X)}\delta(\D^r-\D(X,R)) \ , \\
F_g[T,\g;R,\D^r] &= -T \log \int dX  e^{-\b V_\g(X)}\delta(\D^r-\D(X,R)) \equiv -T \log Z_g \ , \\
  Z_m & = \int dR^1 \cdots dR^m \, e^{-\b_g \sum_{a=1}^m V(R^a)} \ , \\
F_g[T,\g;\D^r] &= \overline{F_g[T,\g;R,\D^r] }
 \equiv
 \frac{1}{Z_m} \int dR^1 \cdots dR^m \, e^{-\b_g \sum_{a=1}^m V(R^a)} F_g[T,\g;R^1,\D^r]
  \ .\label{eq:FP_definition}
\end{split}
\eeq
Note that the above definition of $F_g[T,\g;R,\D^r]$ is slightly different from the one we have given in the main text, because we replaced the Heaviside step function
with a Dirac delta function.
With this choice, $F_g[T,\g;\D^r]$ is the averaged (over glassy states) large deviation function of the mean square displacement $\D^r$: it gives the thermodynamic
weight of all the configurations of $X$ that are at distance $\D^r$ from $R$.
Above the dynamical packing fraction $\f_{\rm d}$ where the first glassy states appear,
the free energy $F_g[T,\g;\D^r]$ develops a minimum at a finite value of $\D^r$, 
signaling the presence of metastable states. 
The intuitive reason for the presence of a secondary minimum in the glass phase
is the following. At very small $\D^r$, there are few configurations, 
so the weight is small and $F_g[T,\g;\D^r]$.
Upon increasing $\D^r$, the weight increases as one explores larger portions of the glass basin around $R$, and $F_g[T,\g;\D^r]$ decreases.
However, if $\D^r$ is increased beyond the size of the glass basins, then the configurations at distance $\D^r$ are on the barriers that surround the 
glassy state, therefore the weight is small and $F_g[T,\g;\D^r]$ increases again. Only at much larger $\D^r$, when all the configurations corresponding
to other glass basins are included, the entropic contribution of all the glass basins makes $F_g[T,\g;\D^r]$ smaller. See~\cite{FP95,BFP97} for a more
detailed discussion, and~\cite{MP00} for a detailed discussion of how to construct $F_g[T,\g;\D^r]$ by adding a coupling to the system, and for a generalization
to this discussion to finite $d$ by taking into account phase coexistence.

Minimizing the constrained free energy $F_g[T,\g;\D^r]$ with respect to $\D^r$ 
is thus the way to obtain the properties of the typical metastable states selected 
by the reference configuration $R$ once followed under an external perturbation \cite{FP95, BFP97, MP99, MP00}.
At the minimum, the value of $\D^r$ corresponds to the configurations that have larger Boltzmann weight in the glass basin,
and $F_g[T,\g;\D^r]$ gives the corresponding weight.
The weight of configurations corresponding to smaller $\D^r$ is exponentially suppressed when $d \to\io$. 
Therefore in the following we assume that $\D^r$ is determined by the minimization of the free energy.
Note that if we define the constrained free energy with a ``soft'' constraint (a Heaviside $\theta$ function, as in the main text) 
instead of a ``hard'' constraint (a Dirac $\d$ function), 
we will get the same value for the saddle point solution for the mean square displacement and the same properties for the metastable states 
because the configurations that we are adding by considering a soft constraint have an exponentially suppressed weight.
This is why we have chosen to use the $\th$ function in the main text, because it is better for illustrative purposes.

We can use the replica trick to compute the logarithm. If we define
\beq\begin{split}
-\b N F_{\rm FP}
 &=  \log \int dR^1 \cdots dR^m dX^{1} \cdots dX^{s} e^{-\b_g \sum_{a=1}^m V(R^a) - \b \sum_{b=1}^{s} V_\g(X^b)} \\
&= \log \int dX^1\cdots dX^m e^{-\b_g \sum_{a=1}^m V(X^a)} (Z_g)^s = \log[ Z_m \overline{(Z_g)^s} ] \ ,
\end{split}\eeq
then we have, at leading order for small $s$
\beq\begin{split}
-\b N F_{\rm FP} &= \log \left[ Z_m  \overline{ (Z_g)^s } \right] \sim  \log \left[ Z_m  (1 + s \overline{ \log(Z_g)} + O(s^2) ) \right] = 
\log Z_m + s \, \overline{ \log Z_g } + O(s^2)  \\ 
&= -\b F_m - s \b F_g[T,\g] + O(s^2)
\ .
\end{split}\eeq
Therefore we have to
compute the free energy of $m+s$ replicas; $m$ ``reference'' ones and $s$ ``constrained'' ones, that are at different temperature or density.
Then we have to send $s\to 0$; the leading order gives the Monasson replicated free energy $F_m$~\cite{Mo95}, 
while the linear order in $s$ gives the Franz-Parisi free energy $F_g[T,\g]$~\cite{FP95}.
In this paper we will only consider the case $m\to 1$, in which $F_m$ coincides with the liquid free energy.

In the following we consider a system of hard spheres in $d\to\io$, hence temperature plays no role and density (or packing fraction) 
is the only relevant control parameter. Furthermore, the energy is zero, therefore the free energy contains only the entropic term $-\b F = s$.
For technical reasons, 
it is convenient to fix the packing fraction through the sphere diameters, while assuming that the number density is constant,
as in the Lubachevsky-Stillinger algorithm~\cite{LS90}.
We consider that the $m$ reference replicas have diameter $D_g$ and packing fraction $\f_g$,
while the $s$ constrained replicas have the same number density but $D = D_g ( 1 + \h/d)$. 
Following~\cite{PZ10,KPZ12} we also define a rescaled packing fraction $\wh \f = 2^d \f/d$ that has a finite limit when $d\to\io$.
Note that the packing fraction of the constrained replicas is therefore
$\f = \f_g (D/D_g)^d \sim \f_g e^\h$ and similarly
$\wh\f = \wh\f_g e^\h$.

Following~\cite{YM10,Yo12,YZ14}, we also apply a shear strain $\g$ to the constrained replicas, which is obtained by deforming linearly
the volume in which the system is contained. 
We call $x'_\m$, with $\mu=1,\cdots,d$, the coordinates in the original reference frame, in which the shear strain is applied.
In this frame, the cubic volume is deformed because of shear strain.
To remove this undesirable feature, we introduce new coordinates $x_\m$ of a ``strained'' frame in which the volume
is brought back to a cubic shape.
If the strain is applied along direction $\mu=2$, then
all the coordinates are unchanged, $x_{\mu} = x_{\mu}'$, except the first
one which is changed according to
\beq
x_{1}' = x_{1} + \g x_{2} \ , \hskip30pt  x_{1} = x_{1}' - \g x_{2}' \ .
\eeq
Let us call $S(\g)$ the matrix such that $x' = S(\g) x$.
In the original frame (where the volume is deformed by strain), two particles of the slave replica
interact with the potential $v(|x'-y'|)$.
If we change variable to the strained frame (where the volume is not deformed), the interaction is 
\beq
v_{\g}(x-y) = v(|S(\g)(x-y)|) \ . 
\eeq
An important remark is that $\det S(\gamma) =1$ meaning 
that the simple strain defined above does not
change the volume and thus the average density $\rho=N/V$ of the system.

In summary, if we consider following a glass state under a compression and a strain, we have to compute the Franz-Parisi potential
where the constrained replicas have a diameter $D = D_g (1 + \h/d)$ 
and interact with a potential $V_\g(X) = \sum_{i<j} v_\g(x_i - x_j)$. The control parameter of the reference replica is their density $\f_g$,
while the control parameters of the constrained replicas are compression rate $\h = \log(\f/\f_g)$ and shear strain $\g$.
The replicated entropy of this system can be computed through a generalization of the methods of Refs.~\cite{KPZ12,KPUZ13},
which we present below.

\subsection{Replicated entropy}

The replicated entropy of the system for a generic replica structure has been derived in~\cite{KPUZ13}:
\beq\label{eq:gauss_r}
s[\hat \a] = 1 - \log\r + d \log (m+s) + \frac{(m+s-1)d}{2} \log(2 \pi e D_g^2/d^2) + \frac{d}2 \log \det(\hat \a^{m+s,m+s}) -  \frac{d}2 \wh \f_g \,
 \FF\left( 2 \hat \a \right) \ ,
\eeq
where $\hat\a = d \la u_a \cdot u_b \ra/D_g^2$ is a $(m+s)\times(m+s)$ symmetric matrix and $\hat \a^{a,a}$ is the matrix obtained from $\hat\a$
by deleting the $a$-th row and column.
The matrix $\hat\a$ encodes the fluctuations of the replica displacements $u_a$ around the center of mass of all replicas. Because $\sum_a u_a=0$,
the sum of each row and column of $\hat\a$ is equal to zero, i.e. $\hat\a$ is a {\it Laplacian} matrix. 
Here we used $D_g$ as the unit of length and for this reason $D_g$ and
$\wh\f_g$ appear in Eq.~\eqref{eq:gauss_r}.
We call the last term in Eq.~\eqref{eq:gauss_r} the ``interaction term'', while all the rest will be called the ``entropic term''.

Given the replica structure of the problem, the simplest replica symmetric (RS) ansatz for the matrix $\hat\a$ is
\beq\label{vtot}
\begin{split}
\hat \a = & 
\begin{bmatrix}
\d^g   & -\a^g   & \cdots            & -\a^g   & -\c   & \multicolumn{4}{c}{  \cdots }          & -\c \\
-\a^g   & \d^g   & \cdots            & -\a^g   &        & \multicolumn{4}{c}{         }          &      \\
\vdots & \ddots & \ddots            & \vdots & \vdots & \multicolumn{4}{c}{         }          & \vdots \\
-\a^g   & \cdots & -\a^g              & \d^g   & -\c   & \multicolumn{4}{c}{  \cdots }          & -\c \\
-\c   & \multicolumn{2}{c}{\cdots} & -\c   & \d   & -\a   & \multicolumn{3}{c} { \cdots } & -\a \\
       &      &                     &        & -\a   & \d   & \multicolumn{3}{c} { \cdots } & -\a \\
\vdots &      &                     & \vdots & \vdots & \vdots &      & \ddots &               & \vdots \\
       &      &                     &        & -\a   & \multicolumn{3}{c} { \cdots } & \d   & -\a \\
-\c   & \multicolumn{2}{c}{\cdots} & -\c   & -\a   & \multicolumn{3}{c} { \cdots } & -\a   & \d \\
\end{bmatrix} \\
\\
&\d^g  = (m-1) \a^g + s \chi \\
&\d  = (s-1) \a + m \chi \\
\end{split}\eeq

Note that the mean square displacements between the replicas are $\D_{ab} = d\la (u_a - u_b)^2\ra/D_g^2=
\a_{aa} + \a_{bb} - 2 \a_{ab}$ (we scaled $\hat\D$ by $D_g$ because we use $\D_g$ as the unit of length)
and therefore the matrix $\hat\D$ has the form
\beq\label{DeltaRS}
\hat \D = 
\begin{bmatrix}
0  & \Dg   & \cdots            & \Dg   & \D^r   & \multicolumn{4}{c}{  \cdots }          & \D^r \\
\Dg   & 0   & \cdots            & \Dg   &        & \multicolumn{4}{c}{         }          &      \\
\vdots & \ddots & \ddots            & \vdots & \vdots & \multicolumn{4}{c}{         }          & \vdots \\
\Dg   & \cdots & \Dg              & 0   & \D^r   & \multicolumn{4}{c}{  \cdots }          & \D^r \\
\D^r   & \multicolumn{2}{c}{\cdots} & \D^r   & 0  & \De   & \multicolumn{3}{c} { \cdots } & \De \\
       &      &                     &        & \De   & 0   & \multicolumn{3}{c} { \cdots } & \De \\
\vdots &      &                     & \vdots & \vdots & \vdots &      & \ddots &               & \vdots \\
       &      &                     &        & \De   & \multicolumn{3}{c} { \cdots } & 0   & \De \\
\D^r   & \multicolumn{2}{c}{\cdots} & \D^r   & \De   & \multicolumn{3}{c} { \cdots } & \De   & 0\\
\end{bmatrix} 
\eeq
with
\beq\label{Deltaer}
\begin{split}
&\Dg = 2 (\d^g + \a^g) = 2(m\a^g+ s\chi) \ , \\
&\De = 2 (\d + \a) = 2 (s \a + m\c) \, \\
&\D^r = \d^g + \d + 2 \c =(m-1)\a^g + (s-1) \a + (m+s+2) \c \ , \\
&\D^f = 2 \D^r - \D^g -\D = 2(2\c - \a^g -\a) \ ,
\end{split}\eeq
where $\D^g$ is internal to the block of $m$ replicas, $\De$ to the $s$ replicas, and $\D^r$ is the relative
displacement between the $m$-type and $s$-type replicas. Finally, we introduced $\D^f$ which measures
the additional fluctuations between the $m$ and $s$-type replicas.
The entropy~\eqref{eq:gauss_r} must be maximized with respect to $\Dg, \De, \D^r$.

\subsection{The entropic term}

We want to compute $\det \hat \a^{m+s,m+s}$, where we recall that 
$\hat \a^{a,a}$ is the matrix obtained from $\hat\a$
by deleting the $a$-th row and column, i.e. it is the $(a,a)$-cofactor of $\hat\a$.
Begin Laplacian, $\hat \a$ has a vanishing determinant. 
Also, the ``Kirchhoff's matrix tree theorem'' states that for Laplacian matrices,
all the cofactors are equal, hence $\det \hat\a^{a,a}$ is independent of $a$.
Therefore, if $\mathbf{1}$ is the identity in $m+s$ dimensions, we have
\beq\label{eq:detmm}
\det(\hat\a + \e \mathbf{1}) = \det\hat\a + \e \sum_{a=1}^{m+s} \det\hat\a^{a,a} + O(\e^2)
\ \ \ 
\Rightarrow 
\ \ \ 
\det\hat\a^{a,a} = \lim_{\e\to 0} \frac{1}{\e(m+s)} \det(\hat\a + \e \mathbf{1}) \ .
\eeq
We then define $\hat \b(\e)=\hat \a + \e \mathbf{1}$ and we note that
\beq\label{eq:beta}
\hat \b(\e)=\begin{pmatrix}
A & B\\
B^T & D
\end{pmatrix}
\eeq
where
$A$ is a $m\times m$ matrix with components
$A_{ab}=(\d^g+\a^g + \e) \d_{ab}-\a^g$,
$D$ is a $s\times s$ matrix 
with $D_{ab}=(\d+\a + \e)\delta_{ab}-\a$, and
$B$ is a $m\times s$ matrix with $B_{ab} = \chi$. 

We can use the following general formula
\beq\label{eq:betae}
\det \hat\b(\e)=(\det A)\det(D-B^TA^{-1}B) \ ,
\eeq
recalling that a $m\times m$ matrix $M_{ab} = M_1 \d_{ab} + M_2$ has determinant $\det M =  M_1^{m-1} (M_1 + m M_2)$ and
its inverse is $M^{-1}_{ab} = (M^{-1})_1 \d_{ab} + (M^{-1})_2$ with 
\beq\label{eq:GM}
\begin{split}
(M^{-1})_1 &= \frac1{M_1} \ , \\
(M^{-1})_2 &= -\frac{M_2}{M_1 ( M_1 + m M_2 )} \ .
\end{split}\eeq
The matrix $A^{-1}$ has this form, with
\beq\label{eq:A}
\begin{split}
(A^{-1})_1 &= \frac{1 }{ \alpha^g +\delta^g +\e } \ , \\
(A^{-1})_2&= \frac{\alpha^g }{(\alpha^g (1- m) + \delta^g  +\e ) (\alpha^g +\delta^g +\e )} \ , \\
\det A &= \left(\d^g +\a^g + \e\right)^{m-1}\left(\d^g+(1-m)\a^g + \e\right) \ .
\end{split}
\eeq
The matrix $\Omega = D - B^T A^{-1} B$ has the same form with
\beq\label{eq:Omega}
\begin{split}
\Omega_1 &= \delta + \e + \a \ , \\
\Omega_2 &= -\a- \c^2 [ m (A^{-1})_1 + m^2 (A^{-1}_2) ] \ , \\
\det \Omega &= (\d + \a + \e)^{s-1} \{ \d + \a (1-s) + \ee -s \c^2[ m (A^{-1})_1 + m^2 (A^{-1}_2) ] \} \ .
\end{split}\eeq
Using Eqs.~\eqref{eq:A}, \eqref{eq:Omega}, \eqref{eq:betae} and \eqref{eq:detmm}, we obtain the final result
\beq\label{entropic_final}
\begin{split}
\det \hat \a^{(m+s,m+s)}&= \c ( m \a^g + s \c)^{m-1} ( s\a + m \c)^{s-1}  
\ .
\end{split}
\eeq

\subsection{The interaction term}

Here we compute the interaction function $\FF(2\hat\a)$. 
This function has been computed in~\cite{KPZ12}, but only for $\h=0$ and $\g=0$. Here we need
to generalize the calculation to non-zero perturbations.

\subsubsection{General expression of the replicated Mayer function}

We follow closely the derivation of~\cite{KPZ12} which has been generalized in~\cite{YZ14}
to the presence of a strain.
The replicated Mayer function is
\beq\begin{split}
\ol f(\bar u ) &= \int dX \left\{  -1 + \prod_{a=1}^m \theta(|X + u_a | - D_g)\prod_{b=m+1}^{m+s} \theta(|S(\g)(X + u_b) | - D) \right\} \\
&= - \int dX \,  \theta\left( \max_{a=1,m+s} \{ D_a -  | S(\g_a) ( X + u_a ) | \} \right) \ ,
\end{split}\eeq
where we introduced $D_a = D_g ( 1 + \h_a/d)$ with $\h_a = \g_a=0$ for $1\leq a \leq m$, and $\h_a = \h$ and $\g_a = \g$ for
$m+1 \leq a \leq m+s$.

The $u_a$ are $m+s$ vectors in $d$ dimensions and define a hyperplane in the $d$-dimensional space.
It is then reasonable to assume that this $(m+s)$-dimensional plane
is orthogonal to the strain
directions $\mu=1,2$ with probability going to 1 for $d\to\io \gg m+s$.
Hence, the vector $X$ can be decomposed in a two dimensional
vector $\{X_1,X_2\}$ parallel to the strain plane, 
a $(d-m-s-2)$-component
vector $X_\perp$, orthogonal to the plane $\mu=1,2$ and to the plane defined by $u_a$,
and a $m+s$-component vector $X_\parallel$ parallel to that plane.
Defining $\Omega_d$ as the $d$-dimensional solid angle and recalling that $V_d=\Omega_d/d$,
and following the same steps as in~\cite[Sec.~5]{KPZ12},
we have, calling $k = m+s$
\beq\begin{split}
\ol f(\bar u ) 
&= -\int dX_1 dX_2 \ d^{k}X_\parallel \ d^{d-k-2}X_\perp \ 
\theta\left( \max_a \{ D_a^2 - (X_1 + \g_a X_2 )^2 - X_2^2 - |X_\parallel + u_a |^2 - | X_\perp |^2 \} \right) \\
&= - \Omega_{d-k-2} \int  dX_1 dX_2 \ d^{k}X_\parallel \int_0^\io dx \, x^{d-k-3} \,  
\theta\left( \max_a \{ D_a^2 - x^2 -  (X_1 + \g_a X_2 )^2 - X_2^2 - | X_\parallel + u_a |^2 \} \right) \\
&= - \Omega_{d-k-2} \int  dX_1 dX_2 \ d^{k}X_\parallel \int_0^{ \sqrt{\max_a \{ D_a^2 -  (X_1 + \g_a X_2 )^2 - X_2^2 - | X_\parallel + u_a |^2 \}} }  dx \, x^{d-k-3} \\
& = - V_{d-k-2}  \int  dX_1 dX_2 \ d^{k}X_\parallel \, \Theta_{d-k-2}\left(\max_a \{ D_a^2 -  (X_1 + \g_a X_2 )^2 - X_2^2 - | X_\parallel + u_a |^2 \} \right)
\end{split}\eeq
where we defined the function
$\Th_{p}(x) = x^{p/2} \th(x)$.

It has been shown in~\cite{KPZ12} that
the region where $\overline{f}(\bar u)$ has a non-trivial dependence on the $u_a$ is 
where $u_a \sim 1/\sqrt{d}$.
Here we use $D_g$ as the unit of length, hence we define
$u_a = x_a D_g/\sqrt{d}$, $X_{1,2} = \z_{1,2} D_g/\sqrt{d}$
and $X_\parallel = \ee D_g/\sqrt{d}$. 
Using that $\lim_{n\to\io} \Th_n(1+y/n) = e^{y/2}$, 
and that for large $d$ and finite $k$ we have $V_{d-k}/V_d \sim d^{k/2} / (2\pi)^{k/2}$,
we have
\beq\begin{split}
\ol f(\bar u ) 
& = - \frac{V_{d-k-2}}{V_d} \frac{V_d D_g^{d}}{d^{(k+2)/2}} \int  d\z_1 d\z_2  d^{k}\ee \, 
\Theta_{d-k-2}\left(1  - \frac1d \min_a\{ -2\h_a + (\z_1 + \g_a \z_2 )^2 + \z_2^2 + | \ee + x_a |^2 \} \right) \\
& \sim - V_d D_g^d  \int \frac{ d\z_1 d\z_2  d^{k}\ee }{(2\pi)^{(k+2)/2}} 
\, e^{  - \frac12 \min_a \{ -2 \h_a + (\z_1 + \g_a \z_2 )^2 + \z_2^2 + | \ee + x_a |^2 \}  } 
\equiv - V_d D_g^d \FF(\bar x )
\ ,
\end{split}\eeq
where the function $\FF$ has been introduced following~\cite{KPZ12,KPUZ13}.

We can then follow the same steps as in~\cite[Sec.V C]{KPUZ13} and in~\cite{YZ14} to obtain
\beq
\begin{split}
\mathcal F(\bar x) &= 
\int \frac{ d\z_1 d\z_2  d^{k}\ee }{(2\pi)^{(k+2)/2}} \, e^{  - \frac12 \min_a \{ -2 \h_a + (\z_1 + \g_a \z_2 )^2 + \z_2^2 + | \ee + x_a |^2 \}  } \\
&=
\int \frac{ d\z_1 d\z_2  d^{k}\ee }{(2\pi)^{(k+2)/2}}
\lim_{n\to 0} \left( \sum_{a=1}^k e^{-\frac1{2n} [  -2 \h_a +(\z_1 + \g_a \z_2 )^2 + \z_2^2 + | \ee + x_a |^2 ] }
\right)^n
\\
&=\lim_{n\to 0}\sum_{n_1,\ldots, n_k; \sum_{a=1}^kn_a=n}\frac{n!}{n_1!\ldots n_k!}
\int \frac{ d\z_1 d\z_2  d^{k}\ee }{(2\pi)^{(k+2)/2}}
e^{ - \sum_a \frac{n_a}{2n} [ -2 \h_a  + (\z_1 + \g_a \z_2 )^2 + \z_2^2 + | \ee + x_a |^2 ]     } \\
\\
&=\lim_{n\to 0}\sum_{n_1,\ldots, n_k; \sum_{a=1}^kn_a=n}\frac{n!}{n_1!\ldots n_k!}
e^{ \sum_{a=1}^k \frac{n_a}{n} \h_a
-\frac 12 \sum_{a=1}^k \frac{n_a}{n}|x_a|^2+\frac{1}{2} \sum_{a,b}^{1,k}\frac{n_an_b}{n^2}x_a\cdot x_b } 
\int \frac{ d\z_1 d\z_2 }{2\pi}
e^{ - \sum_a \frac{n_a}{2n} [ (\z_1 + \g_a \z_2 )^2 + \z_2^2  ]     }
\\
&=\lim_{n\to 0}\sum_{n_1,\ldots, n_k; \sum_{a=1}^kn_a=n}\frac{n!}{n_1!\ldots n_k!}e^{ \sum_{a=1}^k \frac{n_a}{n} \h_a- \frac{1}{4} \sum_{a,b}^{1,k}\frac{n_an_b}{n^2} (x_a - x_b)^2 } 
\int \frac{ d\z }{\sqrt{2\pi}}
e^{ - \frac{\z^2}2 \left[ 1 + \frac12 \sum_{ab} \frac{n_a n_b}{n^2} (\g_a - \g_b)^2 \right]    } \ .
\end{split}\eeq
We now introduce the matrix $\hat\D$ of mean square displacements between replicas
\beq
\D_{ab} = (x_a - x_b)^2 = \frac{d}{D_g^2} (u_a - u_b)^2 \ .
\eeq
We should now recall that the Mayer function is evaluated in $\bar u - \bar v$,
hence after rescaling the function $\FF$ is evaluated in $\bar x - \bar y$.
For $d\to\io$, the interaction term is dominated by a saddle point on $\bar u$ and $\bar v$, such that
$(x_a - x_b)^2 = (y_a - y_b)^2 = \D_{ab}$ and $x_a \cdot y_b =0$~\cite{KPZ12,KPUZ13,CKPUZ13}, hence
$(x_a - y_a -x_b + y_b)^2 = (x_a - x_b)^2 + (y_a - y_b)^2 = 2 \D_{ab}$. This is also why the function $\FF$ is evaluated in $2\hat\a$ in Eq.~\eqref{eq:gauss_r}. 
The contribution of the interaction term to the free energy~\eqref{eq:gauss_r} is~\cite{KPZ12}
\beq
\frac12 \frac{N}V \ol f(\bar u - \bar v) = - \frac{N V_d D_g^d}{2V} \FF(\bar x - \bar y) = - 2^{d-1} \f_g \FF(2\hat\a) = - \frac{d \wh\f_g}2 \FF(2\hat\a) \ .
\eeq
With an abuse of notation, we now call $\FF(\hat\D) = \FF(2\hat\a)$.

We therefore
obtain at the saddle point
\beq\label{FFreplica}
\begin{split}
\mathcal F(\hat\D) 
&=\lim_{n\to 0}\sum_{n_1,\ldots, n_k; \sum_{a=1}^k n_a=n}\frac{n!}{n_1!\ldots n_k!}
e^{\sum_{a=1}^k \frac{n_a}{n} \h_a- \frac{1}{2} \sum_{a,b}^{1,k}\frac{n_an_b}{n^2} \D_{ab} } 
\int \frac{ d\z }{\sqrt{2\pi}}
e^{ - \frac{\z^2}2 \left[ 1 + \frac12 \sum_{ab} \frac{n_a n_b}{n^2} (\g_a - \g_b)^2 \right]    } \\
&=\int \frac{ d\z }{\sqrt{2\pi}}
e^{ - \frac{\z^2}2 }
\lim_{n\to 0}\sum_{n_1,\ldots, n_k; \sum_{a=1}^k n_a=n}\frac{n!}{n_1!\ldots n_k!}
e^{\sum_{a=1}^k \frac{n_a}{n} \h_a- \frac{1}{2} \sum_{a,b}^{1,k}\frac{n_an_b}{n^2} \left( \D_{ab} + \frac{\z^2}{2} (\g_a - \g_b)^2 \right) } 
 \\
&=\int \frac{ d\z }{\sqrt{2\pi}}
e^{ - \frac{\z^2}2 }
\FF_0\left( \D_{ab} + \frac{\z^2}{2} (\g_a - \g_b)^2 \right) \ , 
\end{split}\eeq
where $\FF_0(\hat\D)$ is the interaction function in absence of strain and is given by
\beq\label{eq:FF0binomial}
\FF_0( \hat\D )  = \lim_{n\to 0}\sum_{n_1,\ldots, n_k; \sum_{a=1}^k n_a=n}\frac{n!}{n_1!\ldots n_k!}
e^{\sum_{a=1}^k \frac{n_a}{n} \h_a- \frac{1}{2} \sum_{a,b}^{1,k}\frac{n_an_b}{n^2}  \D_{ab}  } \ .
\eeq

\subsubsection{Computation of the interaction term for a RS displacement matrix}

We now compute the function $\FF_0(\hat\D)$.
for the replica structure encoded by the matrix~\eqref{vtot}.
Defining $\Si_m = \sum_{a=1}^m \frac{n_a}n$ and $\Si_s = \sum_{a=m+1}^{m+s} \frac{n_a}n$, keeping in mind that
$\Si_m + \Si_s = 1$, and recalling that $\h_a = \h$ for $m+1 \leq a \leq m+s$ and $\h_a=0$ otherwise,
we can then write with some manipulations
\beq
\begin{split}
\FF_0(\hat \D) & =\lim_{n\to 0}\sum_{n_1,\ldots, n_{m+s}; \sum_{a=1}^{m+s} n_a=n}\frac{n!}{n_1!\ldots n_{m+s}!} e^{- \left(\frac{\Dg}2 + \frac{\D^f}2 \right) \Si_m - \left( \frac{\De}2- \h \right) \Si_s  + 
 \frac{ \D^f }2 \Si_m^2 
 + \frac\Dg{2} \sum_{a=1}^m \frac{n_a^2}{n^2} + \frac{\De}2 \sum_{a=m+1}^{m+s} \frac{n_a^2}{n^2} } \ . 
\end{split}\eeq

We now introduce Gaussian multipliers to decouple the quadratic terms
and introduce the notation
$\DD \l = \frac{d\l}{\sqrt{2\pi}} e^{-\l^2/2}$.
Note that $\Dg \geq 0$ and $\De \geq 0$. {\it Under the assumption that $2 \D^f \geq 0$} (to be discussed
later on),
we get
\beq\begin{split}
\FF_0(\hat \D)  & = 
\int \DD \l_a \DD \m \lim_{n\to 0}\left[ 
\sum_{a=1}^m
e^{-\frac1{n} \left(\frac{\Dg}2 + \frac{\D^f}2 + \l_a \sqrt{\Dg} + \m \sqrt{ \D^f} \right)} 
+\sum_{a=m+1}^{m+s} e^{-\frac1{n} \left( \frac{\De}2- \h + \l_a \sqrt{\De}  \right)}
\right]^n \\
& = 
\int \DD \l_a \DD \m \,
e^{
-\min\left\{ 
\frac{\Dg}2 + \frac{\D^f}2  + (\min_{a\leq m} \l_a) \sqrt{\Dg} + \m \sqrt{\D^f}, \,
\frac{\De}2- \h + (\min_{a>m} \l_a) \sqrt{\De}  
\right\}
}
\end{split}\eeq
Now we use that for any function $f(x)$,
\beq\begin{split}
& \int \DD \l_a f\left(\min_{a\leq m} \l_a \right) = 
m \int \DD \l f(\l) \left( \int_\l^\io \DD\l' \right)^{m-1} =
-\int d\l f(\l) \frac{d}{d\l}  \Th\left(  -\frac{\l}{\sqrt{2}} \right)^m \equiv \int  \ol\DD_m[ \l] f(\l) \\
&  \ol\DD_n[ \l] = - d\l \frac{d}{d\l} \Th\left(  -\frac{\l}{\sqrt{2}} \right)^n \\
\end{split}\eeq
and we obtain
\beq\begin{split}
\FF_0(\hat \D)  & = 
\int \ol\DD_m \l \, \ol \DD_s \l' \, \DD \m \,
e^{
-\min\left\{ 
\frac{\Dg}2 + \frac{\D^f}2  + \l \sqrt{\Dg} + \m \sqrt{\D^f}, \,
\frac{\De}2- \h + \l' \sqrt{\De}  
\right\}
}
\end{split}\eeq
The integral over $\mu$ can be done and we obtain
\beq\label{eq:Kdef}
\begin{split}
&K(\l,\l') = \int \DD \m \,
e^{
-\min\left\{ 
\frac{\Dg}2 + \frac{\D^f}2  + \l \sqrt{\Dg} + \m \sqrt{\D^f}, \,
\frac{\De}2- \h + \l' \sqrt{\De}  
\right\}
} \\
& = 
e^{-\frac{\De}2+ \h - \sqrt{\De} \l'} \Th\left(\frac{ \h + \frac{\D^f + \Dg - \De}2 + \sqrt{\Dg} \l - \sqrt{\De} \l' }
{\sqrt{2\D^f}}
\right) + e^{-\Dg/2 - \sqrt{\Dg} \l} \Th\left(\frac{-\h + \frac{\D^f - \Dg + \De}2 - \sqrt{\Dg} \l + \sqrt{\De} \l' }
{\sqrt{2 \D^f }}
\right)
\end{split}\eeq

Now by integrating by parts we can write
\beq\begin{split}
\FF_0(\hat \D)  & = 
\int  d\l \frac{d}{d\l} \left[1 - \Th\left(  -\frac{\l}{\sqrt{2}} \right)^m \right] \, 
\int d\l' \frac{d}{d\l'} \left[ 1 - \Th\left(  -\frac{\l'}{\sqrt{2}} \right)^s \right] 
K(\l,\l') \\
&= \int d\l \frac{d}{d\l} \left[1 - \Th\left(  -\frac{\l}{\sqrt{2}} \right)^m \right]
\left\{
K(\l,\l'=\io) - \int d\l' \left[ 1 - \Th\left(  -\frac{\l'}{\sqrt{2}} \right)^s \right] \frac{\partial K}{\partial \l'}(\l,\l')
\right\} \\
&= \int d\l \frac{d}{d\l} \left[1 - \Th\left(  -\frac{\l}{\sqrt{2}} \right)^m \right]
e^{-\Dg/2 - \sqrt{\Dg} \l}
- \int d\l' \left[ 1 - \Th\left(  -\frac{\l'}{\sqrt{2}} \right)^s \right] 
\int d\l \frac{d}{d\l} \left[1 - \Th\left(  -\frac{\l}{\sqrt{2}} \right)^m \right]
\frac{\partial K}{\partial \l'}(\l,\l') \\
&=
\sqrt{\Dg} \int d\l \left[1 - \Th\left(  -\frac{\l}{\sqrt{2}} \right)^m \right]
e^{-\Dg/2 - \sqrt{\Dg} \l}\\
& - \int d\l' \left[ 1 - \Th\left(  -\frac{\l'}{\sqrt{2}} \right)^s \right] 
\left\{
\frac{\partial K}{\partial \l'}(\l=\io,\l') - 
\int d\l \left[1 - \Th\left(  -\frac{\l}{\sqrt{2}} \right)^m \right]
\frac{\partial^2 K}{\partial \l \partial \l'}(\l,\l')
\right\} \\
&=\sqrt{\Dg} \int d\l \left[1 - \Th\left(  -\frac{\l}{\sqrt{2}} \right)^m \right]
e^{-\Dg/2 - \sqrt{\Dg} \l}\\
& +\sqrt{\De} \int d\l' \left[ 1 - \Th\left(  -\frac{\l'}{\sqrt{2}} \right)^s \right] 
e^{-\De/2 + \h - \sqrt{\De} \l'}
\\
& + \int d\l \, d\l' 
 \left[1 - \Th\left(  -\frac{\l}{\sqrt{2}} \right)^m \right]
\left[ 1 - \Th\left(  -\frac{\l'}{\sqrt{2}} \right)^s \right] 
\frac{\partial^2 K}{\partial \l \partial \l'}(\l,\l') \ .
\end{split}\eeq
We also have
\beq\label{K}
\frac{\partial^2 K}{\partial \l \partial \l'}(\l,\l') = - \sqrt{\Dg \De} \, e^{\h - \De/2 -\sqrt{\De} \l'} \,
\frac{ e^{- \frac1{2 \D^f} \left( -\h + \frac{ \De - \Dg - \D^f}2 - \sqrt{\Dg} \l + \sqrt{\De} \l' \right)^2 }  }{\sqrt{2\pi \D^f}} \ .
\eeq
An important remark is that the function $K$ does not depend explicitly on $m$ and $s$,
therefore the derivatives with respect to $m$ and $s$ can be computed straightforwardly.
Also, using Eq.~\eqref{K} one can write
\beq\begin{split}
\FF_0(\hat \D)  
&=\sqrt{\Dg} \int d\l \left[1 - \Th\left(  -\frac{\l}{\sqrt{2}} \right)^m \right]
e^{-\Dg/2 - \sqrt{\Dg} \l}\\
& + \sqrt{\De} \int d\l' 
\left[ 1 - \Th\left(  -\frac{\l'}{\sqrt{2}} \right)^s \right] 
e^{-\De/2 + \h - \sqrt{\De} \l'} 
  \int d\l \, 
 \Th\left(  -\frac{\l}{\sqrt{2}} \right)^m \,
\sqrt{\Dg} \frac{ e^{- \frac1{2 \D^f} \left(- \h + \frac{ \De - \Dg - \D^f}2 - \sqrt{\Dg} \l + \sqrt{\De} \l' \right)^2  } }{\sqrt{2\pi \D^f}} \ .
\end{split}\eeq
We can also change to variables $y = -\Dg/2 -\sqrt{\Dg} \l$ and $y' = \h -\De/2 -\sqrt{\De} \l'$, and $x= y' - y$. 
Then we have
\beq\label{FF0_final}
\begin{split}
\FF_0(\Dg,\De,\D^f)  
&= \int dy \, e^{y} \, \left\{  1 
- \Th\left(  \frac{y + \Dg/2}{\sqrt{2\Dg}} \right)^m \int dx \,  \Th\left(  \frac{ x+ y - \h +\De/2}{\sqrt{2 \De}} \right)^s 
\frac{ e^{- \frac1{2 \D^f} \left( x - \D^f/2  \right)^2  } }{\sqrt{2\pi \D^f}} \right\} \ .
\end{split}\eeq
From Eq.~\eqref{FFreplica}, recalling that $\g_a = \g$ for $m+1\leq a \leq m+s$ and zero otherwise, we have
\beq\label{FF_final}
\FF(\Dg,\De,\D^f)  =
\int \frac{ d\z }{\sqrt{2\pi}}
e^{ - \frac{\z^2}2 }
\FF_0\left( \Dg,\De,\D^f + \z^2 \g^2 \right) \ .
\eeq

\subsection{Final result for the internal entropy of the planted state}

The final result for the replicated entropy is obtained collecting Eqs.~\eqref{eq:gauss_r}, \eqref{entropic_final} and \eqref{FF0_final}-\eqref{FF_final}.
To obtain the Franz-Parisi entropy,
we have to develop the entropy for small $s$ and take the leading order in $s$.
For $s\to 0$ we obtain the Monasson 1RSB entropy~\cite{PZ10,KPZ12}:
\beq\begin{split}
\lim_{s\to 0} s[\hat \a] = s_{m}(\Dg) & = 1 -\log\r + \frac{d}2 (m-1) + \frac{d}2 \log m + \frac{d}2 (m-1) \log( \pi \Dg/d^2 ) \\& -\frac{d}2 \wh\f_g 
 \int dy \, e^y \, \left[1 - \Th\left(  \frac{y + \Dg/2}{\sqrt{2\Dg}} \right)^m \right] \ .
\end{split}\label{eq:seref}\eeq
These determine, for each $m$ and $\wh\f_g$, the cage radius $\Dg$ of the reference configuration.

The linear order in $s$ gives the internal entropy of the glass state sampled by the constrained replicas (Franz-Parisi entropy):
\beq\label{eq:seint}
\begin{split}
&\lim_{s\to 0} \partial_s \{ s[\hat \a] \} = s_{g}  
 = \frac{d}2 + \frac{d}2 \frac{\Dg  + m \D^f }{m \De} + \frac{d}2 \log(\pi \De/d^2) \\
&+\frac{d \wh\f_g}2
\int \DD\z
  \int  dy \,e^{y} \, \Th\left(  \frac{y + \Dg/2}{\sqrt{2\Dg}} \right)^m \int dx \, \log\left[ \Th\left(  \frac{ x+ y - \h +\De/2}{\sqrt{2 \De}} \right) \right] 
\frac{ e^{- \frac1{2 \D_\g(\z)} \left( x - \D_\g(\z)/2  \right)^2  } }{\sqrt{2\pi \D_\g(\z)}}  \ ,
\end{split}\eeq
where $\D_\g(\z) = \D^f + \z^2 \g^2$ and we recall that $\DD \z = \frac{ d\z }{\sqrt{2\pi}}
e^{ - \frac{\z^2}2 }$.
It will be often convenient to make a change of variable $x' = (x - \D_\g(\z)/2 ) / \sqrt{ \D_\g(\z) }$ in the integral,
which leads to (dropping the prime for convenience):
\beq\label{eq:seint2}
\begin{split}
 s_{g}  
& = \frac{d}2 + \frac{d}2 \frac{\Dg  + m \D^f }{m \De} + \frac{d}2 \log(\pi \De/d^2) \\
&+\frac{d \wh\f_g}2
  \int  dy \,e^{y} \, \Th\left(  \frac{y + \Dg/2}{\sqrt{2\Dg}} \right)^m \int \DD\z \, \DD x \, \log\left[ \Th\left(  \frac{ \sqrt{ \D_\g(\z) } x +\D_\g(\z)/2   + y - \h +\De/2}{\sqrt{2 \De}} \right) \right] 
 \ .
\end{split}\eeq
From this expression of the internal entropy, we can obtain the equations for $\De$ and $\D^f$ and study the behavior
of glass states.

\subsection{Derivation from the Gaussian replica method}
\label{sec:Gaussian}

As a side remark, we note that following the general strategy outlined in~\cite{KPZ12,CKPUZ13}, the 
same results can be also derived in the replica scheme directly from a Gaussian assumption for the cage shape.
The starting point is the expression of the replicated entropy as a functional of the single-molecule density 
$\r(\overline{x})$~\cite{KPZ12,CKPUZ13}.
The appropriate {\it ansatz} that corresponds to the replica structure in Eq.~\eqref{DeltaRS} has the form
\beq\label{Gaussian}
\r(\overline{x}) = \r \int dX dY \g_{\DE^f/2}(X-Y) \left( \prod_{a=1}^m \g_{\DE^g/2}(x_a-X) \right) \left( \prod_{b=m+1}^{m+s} \g_{\DE/2}(x_b-Y) \right) \ ,
\eeq
where $\g_{A}(x)$ is a normalized Gaussian of variance $A$, and the coefficients $\D = d^2 \DE / D_g^2$.
The Gaussian approximation is exact in $d\to\io$~\cite{KPZ12} and
it is useful to derive approximate expressions in finite dimensions~\cite{CKPUZ13}.

\subsection{Stability of the RS solution}

In this section we discuss the stability of the replica symmetric ansatz \eqref{DeltaRS} for the calculation of the Franz-Parisi entropy.
We want to compute the stability matrix of the small fluctuations around the RS solution and from that extract the replicon eigenvalue~\cite{KPUZ13}.
This calculation is very close to the one given in Ref. \cite{KPUZ13} and we will use many of the results reported in that work.

\subsubsection{The structure of the unstable mode}
\label{sec:replicongen}

The general stability analysis of the RS solution can be done on the following lines. 
We have to take the general expression (\ref{eq:gauss_r}) and compute the Hessian matrix 
obtained by varying at the second order the replicated entropy with respect to the full matrix $\hat \a$. We can then compute the Hessian on the RS saddle point.
The task here is complicated by the fact that the entropy (\ref{eq:gauss_r}) is not symmetric under permutation of all replicas. The symmetries are restricted
to arbitrary perturbations of the $m$ replicas and the $s$ replicas separately. Hence the structure of the Hessian matrix is more complicated than the one
studied in~\cite{KPUZ13}. 

However, here we are mostly interested in studying the problem when the $m$ replicas are at equilibrium in the liquid phase, hence $m=1$, and in that
case we already know that the RS solution is stable in the sector of the $m$ replicas~\cite{KPUZ13}. 
Moreover, the $m$ reference replicas evolve dynamically without being influenced 
by the constrained ones.
Hence, on physical grounds, we expect that replica symmetry will be broken in the sector of the $s$ replicas and that the unstable mode in that sector
will have the form of a ``replicon'' mode similar to the one studied in~\cite{KPUZ13}.
In fact, the $s$ replicas have the task to probe the bottom of the glassy basins identified by the reference replicas, 
and they may thus fall in the Gardner phase when the glassy state identified by the $m$ replicas is followed at sufficiently large pressures
or low temperatures.
Based on this reasoning, we conjecture the following form for the unstable mode:
\beq\label{eq:replicon}
\d \hat\D = \left[
\begin{matrix}
\d \Dg (I^{m}_{ab} - \d_{ab}) &    \d \D^r I_{ab}^{m,s} \\
\d \D^r I_{ab}^{s,m}    & \d \D_R \, r_{ab} \\
\end{matrix} \right] \ ,
\eeq
where $\hat I^m$ is a $m\times m$ matrix and $\hat I^{m,s}$ is a $m\times s$ matrix with all elements equal to $1$, 
and $\hat r$ is a $s\times s$ ``replicon'' matrix such that $\sum_{ab} r_{ab}=0$~\cite{KPUZ13,CKPUZ13}.
In other words, we look for fluctuations around the RS matrix \eqref{DeltaRS} where the matrix elements of the $m$
replicas $\D$ and the matrix elements connecting the $m$ and $s$ replicas $\D^r$ are varied uniformly, while in the
$s$ block we break replica symmetry following the replicon mode.

Let us write the variation of the entropy \eqref{eq:gauss_r} around the RS solution, along the unstable mode \eqref{eq:replicon}. We have
\beq
\d s = \frac12 \sum_{a\neq b, c\neq d} M_{ab;cd} \d\D_{ab} \d\D_{cd} + \frac16 \sum_{a\neq b, c\neq d, e\neq f} W_{ab;cd;ef} \d\D_{ab} \d\D_{cd} \d\D_{ef} 
+ \cdots
\ .
\eeq
The mass matrix $M_{ab;cd}$ and the cubic term $W_{ab;cd;ef}$ are derivatives of the entropy 
$s$ (which is replica symmetric) computed in a RS point and
therefore they must stasify certain symmetries which are simple extensions of the ones discussed in~\cite{KPUZ13}.
Let us call $(ab)^m$ a pair of indeces $a\neq b$ that both belong to the $m$ block. Similarly $(ab)^s$ belong to the $s$ block, and 
$(ab)^r$ are such that one index belong to the $m$ block and the other to the $s$ block.
At the quadratic order, we obtain
\beq\begin{split}
\d s &= \frac12 (\d\Dg)^2 \sum_{(ab)^m, (cd)^m} M_{ab;cd} 
+\frac12 (\d\D^r)^2 \sum_{(ab)^r, (cd)^r} M_{ab;cd} 
+ \frac12 \d\D_R^2 \sum_{(ab)^s, (cd)^s} M_{ab;cd} r_{ab} r_{cd} \\
&+ \d\Dg \d\D^r \sum_{(ab)^m,(cd)^r} M_{ab;cd}
+ \d\Dg \d\D_R \sum_{(ab)^m,(cd)^s} M_{ab;cd} r_{cd}
+ \d\D^r \d\D_R \sum_{(ab)^r,(cd)^s} M_{ab;cd} r_{cd} \ .
\end{split}\eeq
It is easy to show that the cross-terms involving the replicon mode vanish. In fact,
the sum $\sum_{(ab)^m} M_{ab;cd}$ must be a constant independent of the choice of indeces $(cd)^s$, which are all
equivalent due to replica symmetry in the $s$-block. Hence $\sum_{(ab)^m,(cd)^s} M_{ab;cd} r_{cd} = const. \sum_{(cd)^s} r_{cd} =0$
because of the zero-sum property of the matrix $\hat r$. The same property applies to the other cross-term.
The quadratic term has therefore the form
\beq
\d s^{(2)} = \frac12 A (\d\Dg)^2 
+\frac12 B (\d\D^r)^2 + C \d\Dg \d\D^r 
+ \frac12 \d\D_R^2 \sum_{(ab)^s, (cd)^s} M_{ab;cd} r_{ab} r_{cd} \ ,
\eeq
and the stability analysis of the replicon mode in the $s$-block can be done independenty of the presence of the $m$ replicas.

A similar reasoning can be applied to the cubic terms. Let us write only the terms that involve the replicon mode:
\beq\begin{split}
\d s^{(3)} &= \frac16 \d\D_R^3 \sum_{(ab)^s, (cd)^s, (ef)^s} W_{ab;cd;ef} r_{ab} r_{cd} r_{ef} +
\frac12 \d\D_R^2 \d\Dg \sum_{(ab)^s, (cd)^s, (ef)^m} W_{ab;cd;ef} r_{ab} r_{cd} \\
&+\frac12 \d\D_R^2 \d\D^r \sum_{(ab)^s, (cd)^s, (ef)^r} W_{ab;cd;ef} r_{ab} r_{cd} +
\d\D_R \d\D^r \d\Dg \sum_{(ab)^s, (cd)^r, (ef)^m} W_{ab;cd;ef} r_{ab} \\ & +
\frac12 \d\D_R (\d\D^r)^2 \sum_{(ab)^s, (cd)^r, (ef)^r} W_{ab;cd;ef} r_{ab}  +
\frac12 \d\D_R (\d\Dg)^2 \sum_{(ab)^s, (cd)^m, (ef)^m} W_{ab;cd;ef} r_{ab} + \text{terms without  } \d\D_R
\end{split}\eeq
Clearly, all terms that are linear in $\d\D_R$ vanish. In fact, for example 
\beq
\sum_{(ab)^s, (cd)^r, (ef)^m} W_{ab;cd;ef} r_{ab} = \sum_{(ab)^s} r_{ab} \sum_{(cd)^r, (ef)^m} W_{ab;cd;ef} = \text{const.} \times  \sum_{(ab)^s} r_{ab} =0 \ ,
\eeq
because once again $\sum_{(cd)^r, (ef)^m} W_{ab;cd;ef}$ must be a constant independent of the choice of $(ab)^s$ which are all equivalent thanks to
replica symmetry in the $s$-block.
Collecting all non-vanishing terms that involve the replicon mode, we obtain
\beq\begin{split}
\d s  &= \frac12 A (\d\Dg)^2 
+\frac12 B (\d\D^r)^2 + C \d\Dg \d\D^r 
+ \frac12 \d\D_R^2 \sum_{(ab)^s, (cd)^s} M_{ab;cd} r_{ab} r_{cd}  \\
 &+ \frac16 \d\D_R^3 \sum_{(ab)^s, (cd)^s, (ef)^s} W_{ab;cd;ef} r_{ab} r_{cd} r_{ef} +
\frac12 \d\D_R^2 \d\Dg \sum_{(ab)^s, (cd)^s, (ef)^m} W_{ab;cd;ef} r_{ab} r_{cd} \\
&+\frac12 \d\D_R^2 \d\D^r \sum_{(ab)^s, (cd)^s, (ef)^r} W_{ab;cd;ef} r_{ab} r_{cd}  \ .
\end{split}\eeq
The resulting entropy should be optimized over $\d\Dg, \d\D^r, \d\D_R$. The above equation clearly shows that for a fixed $\d\D_R$,
the optimization over $\d\Dg, \d\D^r$ given $\d\Dg \sim \d\D^r \sim \d\D_R^2$. Hence we conclude that all the terms that involve 
$\d\Dg$ and $\d\D^r$ are at least of order $\d\D_R^4$ and can be neglected in the linear stability analysis. We finally obtain at the leading
order
\beq
\d s  =  \frac12 \d\D_R^2 \sum_{(ab)^s, (cd)^s} M_{ab;cd} r_{ab} r_{cd} 
+ \frac16 \d\D_R^3 \sum_{(ab)^s, (cd)^s, (ef)^s} W_{ab;cd;ef} r_{ab} r_{cd} r_{ef} 
\eeq
and all the couplings between the $s$-block and the $m$-block disappear. 
This shows that the stability analysis of the replicon mode
can be performed by restricting all the derivatives to the $s$-block, both at the quadratic and cubic orders. The Gardner transition corresponds
to the appearance of a negative mode in the quadratic term for a particular choice of the matrix $r_{ab}$ that corresponds to a 1RSB
structure in the $s$-block, characterized by a Parisi parameter $s_1$, as discussed
in~\cite[Sec.~VII]{CKPUZ13}. The unstable quadratic mode is stabilized by the cubic term leading to a fullRSB phase~\cite{Ri13,CKPUZ13}.
Note that, according to the analysis of~\cite{Ri13,CKPUZ13}, in the ``typical state'' calculation done with $m$ replicas with $m\in[0,1]$ taken as
a free parameter, the fullRSB phase
can only be stabilized if the parameter $m_1 > m$, and this only happens at low enough temperature or large
enough densities, hence the fullRSB phase can only exist at sufficiently low temperatures and high densities~\cite{Ri13,CKPUZ13}. 
However is situation is crucially different here because the state following construction requires $s\to 0$. The perturbative analysis
gives $s_1 = \l(s)$, where $\l(s) > 0$ is the MCT parameter discussed in~\cite{KPUZ13,CKPUZ13}, hence one always have 
$s_1 = \l(s) > s =0$ and the fullRSB phase exist at all temperatures and densities when the RS phase becomes unstable.

In summary, we have shown that
we can define the following stability matrix
\beq
M_{a	\neq b;c\neq d}=\frac 2d \frac{\d^2 s[\hat \a]}{\d \a_{a<b}\d \a_{c<d}}=M_1\left(\frac{\d_{ac}\d_{bd}+\d_{ad}\d_{bc}}{2}\right)+M_2\left(\frac{\d_{ac}+\d_{ad}+\d_{bc}+\d_{bd}}{4}\right)+M_3
\label{stab_matr}
\eeq
where the indices $a,\ b,\ c,\ d$ run between $m+1$ and $m+s$.
The fact that the replica structure of this stability matrix is the one defined in Eq. (\ref{stab_matr}) is due to
replica symmetry under permutation of the $s$ replicas.
When a zero mode appears in this matrix, the replica solution becomes unstable and transform continuously in a fullRSB phase, signaling that the glass state
sampled by the $s$ replicas undergoes a Gardner transition.

In the following, we divide the problem of computing that stability matrix in the part coming from the derivatives of the entropic term and the part relative to the interaction term. 
We will first derive the stability matrix in the case of absence of shear and we will discuss the generalization of the method only at the end.

\subsubsection{Entropic term part of the stability matrix}

We want to compute first the contribution of the entropic term to the stability matrix.
Note that under a variation of $\d\a_{ab}$, we have an identical variation of $\d\a_{ba} = \d\a_{ab}$, 
and the diagonal terms vary by minus the same amount, $\d\a_{aa} = \d\a_{bb} = -\d\a_{ab}$ to
maintain the Laplacian condition of $\hat\a$.
Hence we have
\beq
\frac{\d}{\d\a_{a<b}} = \frac{\d}{\d\a_{ab}} +\frac{\d}{\d\a_{ba}} - \frac{\d}{\d\a_{aa}} - \frac{\d}{\d\a_{bb}} \ .
\eeq
From Eq.~\eqref{eq:detmm}, recalling that $\hat \b(\e)=\hat \a + \e \mathbf{1}$,
we have $\log\det\hat\a^{m+s,m+s} = \log\det \hat\b - \log(\e) - \log(m+s) + O(\e)$, therefore, using (for symmetric matrices)
\beq
\frac{\d}{\d\b_{ab}} \log\det\hat \b = \b^{-1}_{ab} \ , \hskip30pt
\frac{\d^2}{\d\b_{ab} \d \b_{cd}} \log\det\hat \b = \frac{\d\b^{-1}_{ab}}{\d\b_{cd}} = -\b^{-1}_{ac} \b^{-1}_{bd} \ ,
\eeq
we obtain
\beq\label{eq:MEapp}
\begin{split}
M_{ab;cd}^{(E)} &=\frac{\d^2}{\d \a_{a<b}\d \a_{c<d}} \log \det \hat \a^{m+s,m+s}= \lim_{\e\to 0} \frac{\d^2}{\d \b_{a<b}\d \b_{c<d}}\log \det \hat \b \\
 &= \lim_{\e\to0}  \left[  -2 \b^{-1}_{ac}\b^{-1}_{bd} - 2 \b^{-1}_{ad}\b^{-1}_{bc}  + 2 \b^{-1}_{ac}\b^{-1}_{bc}  +2 \b^{-1}_{ad}\b^{-1}_{bd} + 2\b^{-1}_{ac}\b^{-1}_{ad}
+ 2 \b^{-1}_{bc}\b^{-1}_{bd} \right. \\
&  \hskip32pt  \left. - (\b^{-1}_{ac})^2 - (\b^{-1}_{bc})^2 - (\b^{-1}_{ad})^2 -  (\b^{-1}_{bd})^2 \right]
\ .
\end{split}\eeq
Based on the discussion above,
we are only interested in the matrix elements corresponding to $a,b,c,d$ belonging to the block of $s$ replicas.
The matrix $\hat\b$ has the form \eqref{eq:beta}, and using the block-inversion formula, its inverse in the $s$ block
is $\Omega^{-1} = ( D - B A^{-1} B^T )^{-1}$. Hence, for $a,b \in [m+1,m+s]$ we have
$\b^{-1}_{ab} = \Omega^{-1}_{ab} = (\Omega^{-1})_1 \d_{ab} + (\Omega^{-1})_2$ where the coefficients are obtained
from Eq.~\eqref{eq:Omega} and Eq.~\eqref{eq:GM}. In particular we have $ (\Omega^{-1})_1 = 1/(\d+\a+\e) = 1/(\De/2+\e)$.

Plugging this form of $\b^{-1}_{ab}$ in Eq.~\eqref{eq:MEapp}, one can check that all terms involving $(\Omega^{-1})_2$ disappear
(as it should, because this term is divergent when $\e\to 0$), so the correct result is obtained by inserting in Eq.~\eqref{eq:MEapp} the form
$\b^{-1}_{ab} = (2/\D) \d_{ab}$, and we get (recalling that $a\neq b$ and $c\neq d$):
\beq\begin{split}
M_{ab;cd}^{(E)} &= M_1^{(E)}\left(\frac{\d_{ac}\d_{bd}+\d_{ad}\d_{bc}}{2}\right)+M_2^{(E)}\left(\frac{\d_{ac}+\d_{ad}+\d_{bc}+\d_{bd}}{4}\right)+M_3^{(E)} \\
&= - \frac{16}{\D^2} \left(\frac{\d_{ac}\d_{bd}+\d_{ad}\d_{bc}}{2}\right)- \frac{16}{\D^2} \left(\frac{\d_{ac}+\d_{ad}+\d_{bc}+\d_{bd}}{4}\right)
\:.
\end{split}\eeq

\subsubsection{The interaction part term of the stability matrix}

We define
the interaction part of the stability matrix in absence of shear as
\beq
M_{ab;cd}^{(I)}=\left.\frac{\d^2 \mathcal F_0[\hat \y]}{\delta \y_{a<b}\d \y_{c<d}}\right|_{\hat \y=2\hat \a_{RS}}=
M_1^{(I)}\left(\frac{\d_{ac}\d_{bd}+\d_{ad}\d_{bc}}{2}\right)+M_2^{(I)}\left(\frac{\d_{ac}+\d_{ad}+\d_{bc}+\d_{bd}}{4}\right)+M_3^{(I)}
\eeq
so that the expression for the matrix coefficients $M_i$ of the full stability matrix is given by
\beq
M_{i}=M_i^{(E)}-4\wh \varphi M_i^{(I)} \ .
\label{stab_matrix_full}
\eeq
The calculation of the derivatives of the interaction term can be done on the same lines and following the same tricks of \cite{KPUZ13}.
Let us start by writing the general expression for the derivatives using the representation~\eqref{eq:FF0binomial} of the function $\FF_0$. 
We have
\beq
\begin{split}
M_{ab;cd}^{(I)}&=\lim_{n\to 0}\sum_{n_1,\ldots,n_{m+s}:\sum_{a=1}^{m+s} n_a=n}\frac{n!}{n_1!\ldots n_{m+s}!}f(n_a,n_b)f(n_c,n_d)\\
&\times\exp\left[- \left(\frac{\Dg}2 + \frac{\D^f}2 \right) \Si_m - \left( \frac{\De}2- \h \right) \Si_s  + 
 \frac{ \D^f }2 \Si_m^2 
 + \frac\Dg{2} \sum_{a=1}^m \frac{n_a^2}{n^2} + \frac{\De}2 \sum_{a=m+1}^{m+s} \frac{n_a^2}{n^2}\right] \ ,
\end{split}
\eeq
where the function $f$ is defined in~\cite[Eq. (45)]{KPUZ13}. As a variant of~\cite[Eq.(46)]{KPUZ13}
we can introduce the following notation
\beq
\begin{split}
\langle O\rangle=\lim_{n\to 0}\sum_{n_1,\ldots,n_{m+s}:\sum_{a=1}^{m+s} n_a=n}\frac{n!}{n_1!\ldots n_{m+s}!} O 
& \exp\left[- \left(\frac{\Dg}2 + \frac{\D^f}2 \right) \Si_m - \left( \frac{\De}2- \h \right) \Si_s  \right.\\
&\left.+ 
 \frac{ \D^f }2 \Si_m^2 
 + \frac\Dg{2} \sum_{a=1}^m \frac{n_a^2}{n^2} + \frac{\De}2 \sum_{a=m+1}^{m+s} \frac{n_a^2}{n^2}\right]\:.
\end{split}
\eeq
The stability matrix can thus be rewritten as~\cite[Eq.(47)]{KPUZ13} where the replica indices run from $m+1$ to $m+s$.
Then we have to compute monomials of the form $\langle n_{a_1}\ldots n_{a_k}/ n^k\rangle$, which can be done in the following way
\beq
\begin{split}
\langle \frac{n_{a_1}\ldots n_{a_k}}{n^k}\rangle &=\lim_{n\to 0}\sum_{n_1,\ldots,n_{m+s}:\sum_{a=1}^{m+s} n_a =n}
\frac{n!}{n_1!\ldots n_{m+s}!}\frac{n_{a_1}\ldots n_{a_k}}{n^k}\int_{-\infty}^\infty\frac{\de \m}{\sqrt{2\pi}}e^{-\mu^2/2}\int_{-\infty}^\infty\left(\prod_{a=1}^{m+s}\frac{\de \l_a}{\sqrt{2\pi}}e^{-\l_a^2/2}\right)\\
&\times \exp\left[-\left(\frac{\Dg}2 + \frac{\D^f}2 \right) \Si_m - \left( \frac{\De}2- \h \right) \Si_s- \m\sqrt{ \D^f } \Si_m 
 - \sqrt \Dg \sum_{a=1}^m \frac{n_a \l_a}{n} - \sqrt{\De} \sum_{a=m+1}^{m+s} \frac{n_a\l_a}{n}\right]=\\
 &=\frac{1}{\De^{k/2}}\int_{-\infty}^\infty \DD \m \int_{-\infty}^\infty\left(\prod_{a=1}^{m} \DD \l_a \right)
 \int_{-\infty}^\infty\left(\prod_{a=m+1}^{m+s}\frac{\de \l_a}{\sqrt{2\pi}}\right)\frac{\partial^k}{\partial \l_{a_1}\ldots \partial \l_{a_k}}e^{-\frac 12\sum_{a=m+1}^{m+s}\l_a^2}\\
&\times e^{-\min\left\{ 
\frac{\Dg}2 + \frac{\D^f}2  + (\min_{a\leq m} \l_a) \sqrt{\Dg} + \m \sqrt{\D^f}, \,
\frac{\De}2- \h + (\min_{a>m} \l_a) \sqrt{\De}  
\right\}
} \ .
\end{split}
\eeq
If $O$ is a function that depends only on the $\l_a$ with $a\in[m+1,m+s]$, then we can define
\beq
\begin{split}
\langle O\rangle &= \int_{-\infty}^\infty \mathcal D\mu\int_{-\infty}^\infty\left(\prod_{a=1}^{m+s}\mathcal D\l_a\right) O \,  e^{-\min\left\{ 
\frac{\Dg}2 + \frac{\D^f}2  + (\min_{a\leq m} \l_a) \sqrt{\Dg} + \m \sqrt{\D^f}, \,
\frac{\De}2- \h + (\min_{a>m} \l_a) \sqrt{\De}  
\right\}
}=\\
&=\int_{-\infty}^\infty\left(\prod_{a=1}^{m+s}\mathcal D\l_a\right) O \, K(\min_{a\leq m} \l_a,\min_{a>m} \l_a)=
\int_{-\infty}^\infty\left(\prod_{a=m+1}^{m+s}\mathcal D\l_a\right) O \ G\left(\min_{a>m} \l_a\right) \ ,
\end{split}
\eeq
where
\beq\label{eq:Gdef}
G(\l')=\int_{-\infty}^\infty \overline{\mathcal D}_m \l \,K(\l,\l') \ .
\eeq
In this way we obtain a generalization of~\cite[Eq.(48)]{KPUZ13}, in the form
\beq
\langle n_{a_1}\ldots n_{a_k}/ n^k\rangle = \frac1{\De^{k/2}} \left\langle
e^{\frac 12\sum_{a=m+1}^{m+s}\l_a^2}\frac{\partial^k}{\partial \l_{a_1}\ldots \partial \l_{a_k}}e^{-\frac 12\sum_{a=m+1}^{m+s}\l_a^2}
\right\rangle \ .
\eeq
The interaction part of the stability matrix is then given by the same reasoning as in~\cite[Eq.(50, 51, 53, 54, 56)]{KPUZ13} where the replica indices must be all shifted by $m$.
The only difference with respect to \cite{KPUZ13} is the definition of the measure used 
to take the average over the variables $\l$s. In fact instead of having~\cite[Eq.(52)]{KPUZ13} we have
\beq
\langle O(\l') \rangle=\int_{-\infty}^\infty \mathcal D \l' \, G(\l') O(\l') = \int_{-\infty}^\infty \mathcal D \l' \bar{\mathcal D}_m \l \,K(\l,\l') \, O(\l') \ .
\eeq
This completes the calculation of the stability matrix.

\subsubsection{The stability matrix in presence of the shear}

The result of the previous section is valid when $\g=0$. Here we generalize the calculation in the case in which also the shear is present. 
The presence of a non vanishing $\g$ is detectable only in the interaction part of the replicated entropy. This means that
the form of the part of the stability matrix coming from the entropic term does not change (even if the actual value of the elements of the matrix changes 
due to the change of the solution of the saddle point equations in presence of the shear) and we need to compute only the new interaction part of the stability matrix.
This can be done using the following line of reasoning.
The interaction part of the stability matrix can be written in this case as
\beq
M_{ab;cd}^{(I,\g)}=\left.\frac{\d^2}{\delta \y_{a<b}\d\y_{c<d}} \int \mathcal D \zeta \mathcal F_0[\D_{ab}+\frac{\z^2}{2}\g^2 \G_{ab}] \right|_{\hat \y=2\hat \a_{RS}}
\eeq
where the matrix $\G_{ab} = 1$ if $a$ belongs to the $m$-block and $b$ to the $s$-block or viceversa, and zero otherwise.
Recalling that $\D_{ab} =\a_{aa}+\a_{bb}-2\a_{ab}$, we have that the relation between $\hat\D$ and $\hat\a$ is linear, therefore
a constant shift of $\hat\D$ induces a constant shift in $\hat\a$, which does not affect the derivatives.
We deduce that
\beq
M_{ab;cd}^{(I,\g)}= \int \mathcal D \zeta \, M_{ab;cd}^{(I,\g=0)}[\D_{ab}+\frac{\z^2}{2}\g^2 \G_{ab}] 
= \int \mathcal D \zeta \, M_{ab;cd}^{(I,\g=0)}[  \Dg,\De,\D^f + \z^2 \g^2 ]
\ .
\eeq
Because $\D^f$ appears only in the kernel $K$, shifting $\D^f$ amounts to change the measure for the average of monomials of $\l$, 
by using a modified kernel
\beq\begin{split}
K^\g(\l,\l') =  \int \mathcal D \zeta & K(\l,\l' ; \Dg,\De,\D^f + \z^2 \g^2) \\ 
 =  \int \mathcal D \zeta & \left[
e^{-\frac{\De}2+ \h - \sqrt{\De} \l'} \Th\left(\frac{ \h + \frac{\D^f + \z^2 \g^2 + \Dg - \De}2 + \sqrt{\Dg} \l - \sqrt{\De} \l' }
{\sqrt{2( \D^f + \z^2 \g^2)}}
\right) \right. \\
& \left. + e^{-\Dg/2 - \sqrt{\Dg} \l} \Th\left(\frac{-\h + \frac{\D^f + \z^2 \g^2 - \Dg + \De}2 - \sqrt{\Dg} \l + \sqrt{\De} \l' }
{\sqrt{2 ( \D^f + \z^2 \g^2) }}
\right)  \right] \ ,
\end{split}\eeq
and the functional expression of the interaction part of the stability matrix has the same form of the $\g=0$ case.

\subsubsection{The replicon eigenvalue}

Following the results of~\cite{KPUZ13}, 
the replicon eigenvalue is given by
\beq
\l_R=\frac{1}{\De^2}\left(-16-8\wh \varphi\L_s(\hat \D)\right) \ ,
\hskip30pt
\L_s(\hat\D) = \langle 
\Th_0(\l)^{s-1}  \LL_s(\l)
\rangle \ ,
\eeq
where the functions $\Th_i(\l)$ are defined in~\cite[Eqs.(42), (43)]{KPUZ13},
the measure over $\l$ is modified according to the previous discussion and
\beq
\begin{split}
\LL_s(\l) = \left[\left(\frac{\Th_1(\l)}{\Th_0(\l)}\right)^2 - \l \frac{\Th_1(\l)}{\Th_0(\l)}\right]
\left[(2 - 2 \l^2) + (s-4) \left(\frac{\Th_1(\l)}{\Th_0(\l)}\right)^2 + 
(6 - s) \l \frac{\Th_1(\l)}{\Th_0(\l)}\right] \ .
\end{split}
\eeq
Although the replicon is perfectly well defined for $s>0$, a subtlety emerges in the limit $s\to 0$ because of some cancellation
between a divergence in the integral over $\l$ and a vanishing prefactor. This is related to the fact that $\LL_s(\l) = s + O(1/\l^2)$,
see Eq.~\cite[60]{KPUZ13}, but plugging $\LL_s(\l) = s$ in the integral makes it divergent as $1/s$. The asymptotic analysis can be
done along the lines of~\cite[Sec.V D]{KPUZ13} and here we consider only the case $m=1$ on which we are mostly interested.
We first write
\beq
\langle 
\Th_0(\l)^{s-1}  \LL_s(\l)
\rangle = \langle 
\Th_0(\l)^{s-1} ( \LL_s(\l) -s )
\rangle +
\langle 
\Th_0(\l)^{s-1}  s
\rangle \ ,
\eeq
and we then observe, following \cite[Sec.V D]{KPUZ13}, that the first term has a finite limit for $s\to 0$ while the second term is singular.
In fact, the second term is dominated by the large $\l$ region, in which, 
using Eqs.~\eqref{eq:Kdef} and \eqref{eq:Gdef}, it is easy to see that $G(\l\to\io) = 1$. Using the asymptotic behavior
of $\Th_0(\l)$~\cite[Eq.(59)]{KPUZ13}, we obtain for $s\to 0$
\beq
\langle 
\Th_0(\l)^{s-1}  s
\rangle \sim s \int_0^\io \DD\l \left( \frac{e^{-\l^2/2}}{\sqrt{2\pi}} \right)^{s-1} \sim 
s \int_0^\io d\l  \, \l \, e^{-s \l^2/2} \to 1 \ .
\eeq
 We thus conclude that
\beq
\lim_{s\to 0}  \langle 
\Th_0(\l)^{s-1}  \LL_s(\l)
\rangle = 1 + \langle 
\Th_0(\l)^{-1}  \LL_0(\l)
\rangle \ ,
\eeq
and the replicon at $s=0$ is
\beq\label{eq:lR}
\De^2 \l_R = -16 -8 \wh\f -8\wh\f  \langle 
\Th_0(\l)^{-1}  \LL_0(\l)
\rangle \ .
\eeq
Stability requires that the RS matrix is a maximum of the entropy, and therefore $\l_R$ is negative in the stable phase.

\section{Computation of physical observables}

Having established the form of the entropy in the replica-symmetric ansatz, and its stability matrix, we can now use these results 
to extract the physical observables.

\subsection{Maximization of the entropy}

The equation for $\Dg$ is obtained by maximizing Eq.~\eqref{eq:seref}. We have
\beq\label{eq:Dg}
0 = \frac{m-1}{m\Dg} + \frac{\wh\f_g}2 \int dy \ e^y \Th\left(\frac{y+\Dg/2}{\sqrt{2\Dg}}\right)^{m-1}
 \frac{e^{-\frac{(y + \Dg/2)^2}{2\Dg}}}{\sqrt{2\pi \Dg}} \left(   \frac12 - \frac{y}\Dg  \right) \ .
\eeq
For a fixed reference density $\wh\f_g$ (and fixed $m$, here we are mostly interested in $m\to 1$), one can solve this equation to obtain $\Dg$.
Then,
the entropy in Eq.~\eqref{eq:seint2} must be maximized with respect to $\De$ and $\Delta^f$ to give the internal entropy of a glass state prepared
at $\wh\f_g$ (the value of $\Dg$ is the equilibrium one corresponding to $\wh\f_g$) and followed at a different state point parametrized by $\h$ and $\g$.
As usual in replica computations, the analytical continuation to $s\to 0$ induces a change in the properties of the entropy, and as a consequence
the solution of the equations for $\De$ and $\Delta^f$ is not a maximum, but rather a saddle-point. However, the correct prescription is not
to look at the concavity of the entropy, but to check
that all the eigenvectors of the stability matrix (and in particular the replicon mode) remain negative, as we discussed above.

The equations for $\De$ and $\Delta^f$ are obtained from the conditions $\frac{\partial s_{g}}{\partial \De} = 0$ and
$\frac{\partial s_{g}}{\partial \D^f} = 0$. Starting from Eq.~\eqref{eq:seint2} and taking the derivatives, we get
\begin{eqnarray}
\label{eq:eqD}
 0 &=& \frac{m\De-\Dg-m\Delta^f}{m \De^2} + \frac{\wh\f_g}2 \int dy \DD x  \DD\z \ e^y\frac{\Th\left(\frac{y+\Dg/2}{\sqrt{2\Dg}}\right)^m}{\Th\left(\frac{\x}{\sqrt{2\De}}\right)}
 \frac{e^{-\frac{\x^2}{2\De}}}{\sqrt{2\pi \De}}  \left( 1 - \frac{\x }{\D}\right) \ , \\
\label{eq:eqDf}
 0 &=& \frac{1}{\De} + \frac{\wh\f_g}2 \int dy \DD x \DD\z \ e^y\frac{\Th\left(\frac{y+\Dg/2}{\sqrt{2\Dg}}\right)^m}{\Th\left(\frac{ \x}{\sqrt{2\De}}\right)}
 \frac{e^{-\frac{\x^2}{2\De}}}{\sqrt{2\pi \De}} \left( 1 + \frac{ x }{\sqrt{\Delta_\gamma(\zeta)}}\right) \ , \\
 \label{eq:xidef}
 \x &=&  \sqrt{ \D_\g(\z) } x +\D_\g(\z)/2 +y-\h+\De/2 \ ,
 \end{eqnarray}
 where $\D_\g(\z) = \D^f + \g^2 \z^2$.
 In some cases, it might be useful to perform an additional change of variables from $y$ to $\xi$.

\subsection{Pressure and shear stress}

Eq.~\eqref{eq:seint2}, computed on the solutions of Eqs.~\eqref{eq:eqD}-\eqref{eq:eqDf}, gives the internal entropy of the glass state as a function of $\eta$ and
$\g$. We can then compute the pressure and the shear stress, that are the derivatives of the entropy with respect to these two parameters. Recall that 
$\De$ and $\Delta^f$ are defined by setting the derivatives of $s_{g}$ with respect to them equal to zero. Then, when we take for example the derivative
of $s_{g}$ with respect to $\g$, it is enough to take the partial derivative.

For a system of hard spheres, the reduced pressure $p = \b P/\r$ is the response of the system to compression and is given by~\cite{PZ10}
\beq
p_g = - \wh\f \frac{\partial s_g}{\partial \wh\f} = - \frac{\partial s_{g}}{\partial \h} \ ,
\eeq
and we get from Eq.~\eqref{eq:seint2}:
\beq
\frac{p_g}{d} = \frac{\wh\f_g}{2}\int dy \DD x \DD \z \ 
e^y\frac{\Th\left(\frac{y+\Dg/2}{\sqrt{2\Dg}}\right)^m}{\Th\left(\frac{\xi}{\sqrt{2\De}}\right)}\frac{e^{-\frac{\xi^2}{2\De}}}{\sqrt{2\pi \De}}
\label{eq:eqpressure}
\eeq
recalling Eq.~\eqref{eq:xidef}.
The $p_g/d$ vs. $\wh\f$ (or $\h = \log(\wh\f/\wh\f_g)$) curve is the equation of state of the corresponding metastable glass.

The response to a shear strain is given by the shear stress, which is defined as~\cite{YZ14}
\beq
\b \s = - \frac{\partial s_g}{\partial \g},
\eeq
and we get from Eq.~\eqref{eq:seint2}:
\beq
\frac{\b \s}{d} = - \g \frac{ \wh\f_g  }2 \int dy \DD x \DD\z \  e^y\frac{\Th\left(\frac{y+\Dg/2}{\sqrt{2\Dg}}\right)^m}{\Th\left(\frac{\x}{\sqrt{2\De}}\right)}
\frac{e^{-\frac{\x^2}{2\De}}}{\sqrt{2\pi \De}} 
\left(1 + \frac{x}{\sqrt{\Delta_\gamma(\zeta)}}\right) \z^2 \ .
\label{eq:eqshear}
\eeq

\subsection{Response to an infinitesimal strain}

It is interesting to consider as a particular case the response of the glass to an infinitesimal strain, $\g\to 0$.
In that case, we have that both $\D_\g(\z) \to \D^f$ and $\x \to \sqrt{ \D^f } x +\D^f/2 +y-\h+\De/2$ become independent of $\z$.
We have thus
\beq\label{eq:mu}
\frac{\b \mu}{d} = \lim_{\g\to 0} \frac{\b \s}{d \g} = - 
\frac{ \wh\f_g  }2 \int dy \DD x  \  e^y\frac{\Th\left(\frac{y+\Dg/2}{\sqrt{2\Dg}}\right)^m}{\Th\left(\frac{\x}{\sqrt{2\De}}\right)}
\frac{e^{-\frac{\x^2}{2\De}}}{\sqrt{2\pi \De}} 
\left(1 + \frac{x}{\sqrt{\Delta^f}}\right) \int \DD\z \ \z^2 = \frac{1}{\De} \ ,
\eeq
where the last equality is obtained by noticing that $\int \DD\z \ \z^2 = 1$ and using Eq.~\eqref{eq:eqDf} in the limit $\g\to 0$, where
again the integral over $\z$ disappears because $\x$ and $\D_\g$ become independent of $\z$. In this way we see that 
$\s/\g \to \mu$, where $\mu$ is the shear modulus of the glass and it is inversely proportional to the cage radius. This provides
an alternative derivation of the results of~\cite{YZ14}.

From Eq.~\eqref{eq:mu} we deduce that for small $\g$ the physical entropy is
\beq
s_g(\h,\g) = s_g(\h,\g=0) - \frac{d}2 \g^2 \frac{1}{\De(\h,\g=0)} + \cdots \ , 
\eeq
where $\De(\h,\g)$ is the solution of Eqs.~\eqref{eq:eqD}-\eqref{eq:eqDf}. Therefore we have
\beq
p_g(\h,\g) = -\frac{\de s_g(\h,\g)}{\de \h} = p_g(\h,\g=0) + 
\frac{d}2 \g^2 \frac{\de}{\de\h} \frac{1}{\De(\h,\g=0)} + \cdots = p_g(\h,\g=0) + \g^2 (\b R(\h)/\r) + \cdots \ ,
\eeq
from which we deduce the expression of the dilatancy $R$ as
\beq\label{eq:R}
\frac{\b R(\h)}{\r} =  \frac{d}2 \frac{\de}{\de\h} \frac{1}{\De(\h,\g=0)} \ .
\eeq

\section{Special limits and approximations}

In this section we discuss some special limits and approximation of the problem, that are very useful to test the full numerical resolution of the equations.

\subsection{Constrained replicas in equilibrium}

When $\h = \g = 0$, the constrained replicas sample the glass basins in the same state point as the reference replicas.
Therefore, it is reasonable to expect that Eqs.~\eqref{eq:eqD} and \eqref{eq:eqDf} admit
$\De = \D^r = \Dg$, hence $\D^f=0$, as a solution. 

To check this, we first analyze Eq.~\eqref{eq:eqD}. For $\h=\g=0$, $\De=\Dg$ and $\D^f=0$, we have $\D_\g(\z)=0$ and $\xi = y + \De/2$.
Therefore the integrand does not depend on $x$ and $\z$ and $\int \DD x \DD \z=1$. Then it is clear that Eq.~\eqref{eq:eqD} becomes
equivalent to Eq.~\eqref{eq:Dg} and is satisfied by our conjectured solution.

The analysis of Eq.~\eqref{eq:eqDf} is slightly more tricky. Setting $\h=0$, $\gamma=0$, $\De = \Dg$ we get, with a change of variable $x' = x \sqrt{\D^f} + \D^f/2$
(and then dropping the prime for simplicity):
\beq
-\frac{2}{\wh\f_g\De} =  \int dy\ e^y\Th\left(\frac{y+\De/2}{\sqrt{2\De}}\right)^m\int dx\, \frac{e^{-\frac{(x-\Delta^f/2)^2}{2\Delta^f}}}{\sqrt{2\pi \Delta^f}} \, \frac{1}{\Th\left(\frac{x+y+\De/2}{\sqrt{2\De}}\right)}\frac{e^{-\frac{(x+y +\De/2)^2}{2\De}}}{\sqrt{2\pi \De}} \left(\frac{x+\Delta^f/2}{\Delta^f}\right) \ .
\label{eq:eqaf1}
\eeq
We now observe that
\beq
 \left(\frac{x+\Delta^f/2}{\Delta^f}\right) \frac{e^{-\frac{(x-\Delta^f/2)^2}{2\Delta^f}}}{\sqrt{2\pi \Delta^f}}
 =  \left( -\frac{d}{dx} + 1 \right) \frac{e^{-\frac{(x-\Delta^f/2)^2}{2\Delta^f}}}{\sqrt{2\pi \Delta^f}} \xrightarrow[\D^f\to 0]{ } -\delta'(x) + \delta(x)
\eeq
where $\d(x)$ is the Dirac delta distribution.
Therefore Eq.~\eqref{eq:eqaf1} becomes
\beq
\begin{split}
& -\frac{2}{\wh\f_g\De} =\int dy\ e^y\Th\left(\frac{y+\De/2}{\sqrt{2\De}}\right)^m\int dx \, \left[ -\delta'(x) + \delta(x) \right]
\frac{1}{\Th\left(\frac{x+y+\De/2}{\sqrt{2\De}}\right)}\frac{e^{-\frac{(x+y +\De/2)^2}{2\De}}}{\sqrt{2\pi \De}} \\
&=  \int dy\ e^y\Th\left(\frac{y+\De/2}{\sqrt{2\De}}\right)^m  \left( \frac{d}{dy} + 1 \right) \left[
\frac{1}{\Th\left(\frac{y+\De/2}{\sqrt{2\De}}\right)}\frac{e^{-\frac{(y +\De/2)^2}{2\De}}}{\sqrt{2\pi \De}}  \right] \\
& = \int dy\, e^y\Th\left(\frac{y+\De/2}{\sqrt{2\De}}\right)^{m-1}  \left( \frac{d}{dy} + 1 \right) \frac{e^{-\frac{(y +\De/2)^2}{2\De}}}{\sqrt{2\pi \De}} 
+ \int dy \, e^y \Th\left(\frac{y+\De/2}{\sqrt{2\De}}\right)^m \left[ \frac{d}{dy} \Th\left(\frac{y+\De/2}{\sqrt{2\De}}\right)^{-1} \right]
\frac{e^{-\frac{(y +\De/2)^2}{2\De}}}{\sqrt{2\pi \De}}  \\
& = \int dy\, e^y\Th\left(\frac{y+\De/2}{\sqrt{2\De}}\right)^{m-1}  \left( \frac12 - \frac{y}{\De} \right) \frac{e^{-\frac{(y +\De/2)^2}{2\De}}}{\sqrt{2\pi \De}} 
-\frac{1}{m-1} \int dy \, e^y \left[ \frac{d}{dy} \Th\left(\frac{y+\De/2}{\sqrt{2\De}}\right)^{m-1} \right]
\frac{e^{-\frac{(y +\De/2)^2}{2\De}}}{\sqrt{2\pi \De}}  \\
& = \frac{m}{m-1} \int dy\, e^y\Th\left(\frac{y+\De/2}{\sqrt{2\De}}\right)^{m-1}  \left(\frac12 - \frac{y}{\De} \right) \frac{e^{-\frac{(y +\De/2)^2}{2\De}}}{\sqrt{2\pi \De}} \ ,
\end{split}
\eeq
where the last line is obtained by integrating by parts the second term in the previous line. This equation is also equivalent to Eq.~\eqref{eq:Dg}.
We thus conclude that for $\h=\gamma=0$, the solutions of Eqs.~\eqref{eq:eqD}-\eqref{eq:eqDf} is $\Delta^f=0$ and $\Dg=\De$ for all $m$.

A particularly interesting case is when $m=1$. In this case the reference replicas are in equilibrium in the liquid phase, and the
constrained replicas therefore sample the glass basins that compose the liquid phase {\it in equilibrium}.
In fact, from Eq.~\eqref{eq:eqpressure} it is quite easy to see that for $m=1$, $\g=\h=0$, $\D^f=0$, $\Dg=\De$, one has 
$\D_\g(\z)=0$ and $\xi = y + \De/2$ and
\beq
p_g = \frac{d \, \wh\f_g}{2}\int dy \, 
e^y \frac{e^{-\frac{(y+\De/2)^2}{2\De}}}{\sqrt{2\pi \De}} = \frac{d\, \wh\f_g}{2} = p_{\rm liq} \ .
\eeq
This shows in particular that the pressure of glass basins {\it merges continuously with the liquid pressure} at $\wh\f_g$.
Moreover, the internal entropy of these {\it equilibrium} glass states is, from Eq.~\eqref{eq:seint2} 
\beq
 s_{g}  
 = d + \frac{d}2 \log(\pi \De/d^2) 
+\frac{d \wh\f_g}2
  \int  dy \,e^{y} \, \Th\left(  \frac{y + \De/2}{\sqrt{2\De}} \right)  \log\left[ \Th\left(  \frac{  y  +\De/2}{\sqrt{2 \De}} \right) \right] 
 \ .
\eeq
This is the same result that has been obtained using the Monasson scheme in~\cite{PZ10}, which is correct because the Monasson and Franz-Parisi
schemes coincide in the equilibrium limit. The difference $\Si = s_{\rm liq} - s_g$ gives the equilibrium complexity of the liquid.

\subsection{Perturbative solution for small $\h$ and $\g$}
\label{subsubsec:pert}

We can also construct a perturbative solution for small $\g,\eta$. 
We start by expanding $s_{g}$ around the point $\D^f=0$ and $\De = \Dg$,
in powers of $\Delta^f$ and $\De-\Dg$
\beq
\d s_{g} = B(\g,\h)\D^f + C(\g,\h) (\De-\Dg) + \frac{1}{2}D(\g,\h)(\D^f)^2 + F(\g,\h)\D^f(\De-\Dg) + \frac{1}{2}E(\g,\h) (\De-\Dg)^2 + \cdots \ ,
\eeq
then maximizing the entropy we obtain
\begin{eqnarray}
\D^f(\g,\h) &=& \frac{F(\g,\h)C(\g,\h)-B(\h,\g)E(\g,\h)}{D(\g,\h)E(\g,\h)-F(\g,\h)^2} \ ,\\
\De(\g,\h) &=& \frac{F(\g,\h)B(\g,\h)-D(\h,\g)C(\g,\h)}{D(\g,\h)E(\g,\h)-F(\g,\h)^2} + \Dg \ .
\end{eqnarray}
These expressions hold only for small $\g$ and $\h$, and they are useful to obtain the behavior of $\D^f$ and $\De$.
In particular, thanks to a cancellation, we find $\D^f \sim \h^2$ for small $\h$ at $\g=0$. This shows that $\D^f$ increases both in
compression and decompression and suggests that we always have $\D^f \geq 0$. This result is important because all the expressions
we derived above are well defined only for $\D^f \geq 0$. They can be analytically continued to $\D^f<0$, but thanks to this perturbative analysis
we see that the analytic continuation is not needed for our purposes.

\subsection{The jamming limit}
\label{sec:jamming}

Another interesting limit is the jamming limit, where the internal pressure of the glass state diverges and
correspondingly its mean square displacement $\De\to 0$~\cite{PZ10}.
To investigate this limit, we specialize to the case $\g=0$ and we consider the limit $\De\to 0$ of Eqs.~\eqref{eq:eqD} and \eqref{eq:eqDf}. 
Using the relation
\beq
\lim_{\mu\to 0} \Th(x/\sqrt{\mu})^\mu = e^{-x^2 \th(-x)} \ ,
\eeq
the leading order of Eq.~\eqref{eq:seint2} is
\beq\label{eq:sgj}
\begin{split}
 s_g  
 & = \frac{d}2 \frac{\Dg  + m \D^f }{m \De}  
-\frac{d \wh\f_g}{4 \De}  \int  dy \,e^{y} \, \Th\left(  \frac{y + \Dg/2}{\sqrt{2\Dg}} \right)^m \int_{-\io}^{\h-y} dx \, (  x+ y - \h )^2  
\frac{ e^{- \frac1{2 \D^f} \left( x - \D^f/2  \right)^2  } }{\sqrt{2\pi \D^f}} + \cdots \\
& = \frac{d}{2 \De} \left\{ \frac{\Dg}m  +  \D^f  
-\frac{ \wh\f_g}{2 }  \int  dy \,e^{y} \, \Th\left(  \frac{y + \Dg/2}{\sqrt{2\Dg}} \right)^m \int_{-\io}^{0} dx \, x^2  
\frac{ e^{- \frac1{2 \D^f} \left( x -y + \h - \D^f/2  \right)^2  } }{\sqrt{2\pi \D^f}} \right\} + \cdots \ .
\end{split}\eeq
Hence, we obtain that $s_g \sim C/\D + \log\D + \cdots$ when $\D \to 0$, where the term $\log\D$ is explicitly present in Eq.~\eqref{eq:seint2}.
Next, we observe that: 
\begin{itemize}
\item The coefficient $C$ should vanish at jamming. This is because $s_g = \De^{-1} C + \log \De + \cdots$, hence the equation for $\De$
is $-\De^{-2} C + \De^{-1} + \cdots = 0$, or equivalently $-C + \De + \cdots =0$,
which shows that when $C  \to 0$, also $\De = C \to 0$. The jamming point is therefore defined by $C\to 0$.
Note by the way that this condition guarantees that $s_g \sim \log \De$ when $\De \to 0$, which is the physically correct behavior of the glass entropy because
particles are localized on a scale $\D$~\cite{PZ10}.
\item The derivative of $C$ with respect to $\D^f$ should also vanish, because it determines the equation for $\D^f$ at leading order in $\De$.
\end{itemize}
The two conditions $C=0$ and $dC/d\D^f=0$ give two equations that determine the values of $\h$ and $\D^f$ at the jamming point, for a fixed glass 
(i.e. at fixed $\wh\f_g, \Dg, m$).
These two equations read
\beq\label{eq:appj}
\begin{split}
0 &= \frac{\Dg}m  +  \D^f  
-\frac{ \wh\f_g}{2 }  \int  dy \,e^{y} \, \Th\left(  \frac{y + \Dg/2}{\sqrt{2\Dg}} \right)^m \int_{-\io}^{0} dx \, x^2  
\frac{ e^{- \frac1{2 \D^f} \left( x -y + \h - \D^f/2  \right)^2  } }{\sqrt{2\pi \D^f}}  \ , \\
0 &= 1 - \frac{ \wh\f_g}{2 } \frac{d}{d\D^f}  \int  dy \,e^{y} \, \Th\left(  \frac{y + \Dg/2}{\sqrt{2\Dg}} \right)^m \int_{-\io}^{0} dx \, x^2  
\frac{ e^{- \frac1{2 \D^f} \left( x -y + \h - \D^f/2  \right)^2  } }{\sqrt{2\pi \D^f}} \  .
\end{split}\eeq
The integrals over $x$ can be done explicitly and we obtain
\beq
\begin{split}
0 &= \frac{\Dg}m  +  \D^f  
+\frac{ \wh\f_g}{2 }  \int  dy \,e^{y + \h -\D^f/2} \, \Th\left(  \frac{y + \h -\D^f/2+ \Dg/2}{\sqrt{2\Dg}} \right)^m 
\left\{
\sqrt{\frac{\D^f}{2\pi}} e^{-\frac{y^2}{2 \D^f}} y - ( \D^f + y^2 ) \Th\left(- \frac{y}{\sqrt{2\D^f}} \right)
\right\},
  \\
0&= 1 + \frac{ \wh\f_g}{2 }  \int  dy \,e^{y + \h -\D^f/2} \, \Th\left(  \frac{y + \h -\D^f/2+ \Dg/2}{\sqrt{2\Dg}} \right)^m
\left\{
\sqrt{\frac{\D^f}{2\pi}} e^{-\frac{y^2}{2 \D^f}}  -( 1+ y ) \Th\left(- \frac{y}{\sqrt{2\D^f}} \right)
\right\}.
\end{split}\eeq
Note that one could get
the same equations by taking directly the $\De \to 0$ limit of Eqs.~\eqref{eq:eqD} and \eqref{eq:eqDf}.
  This system of two equations determines the values of the jamming density $\wh \f_j = \wh \f_g e^{\h_j}$ 
and the corresponding $\Delta^f_j$. 
Note that in general $C \sim |\h - \h_j|$ and therefore $\De = C \sim |\h - \h_j|$ vanishes linearly at jamming.

\subsection{Following states under compression: approximation $\D^f=0$}
\label{sec:appr0}

We will see later that the numerical solution of Eqs.~\eqref{eq:eqD} and \eqref{eq:eqDf} in compression or decompression
(i.e. at $\g=0$ and $\h\neq 0$) shows that $\D^f$ remains quite small. This motivates considering $\D^f=0$ as an approximation
of the exact RS result. The advantage of this approximation is that the numerical solution of the remaining equation for $\De$ is very
easy. Also, this approximation allows one to grasp most of the physics in compression and decompression. Here we specialize to $m=1$ for simplicity.

Under this approximation we have
\beq\label{eq:sedinfDfzero}
 s_g =  \frac{d}2 +
\frac{d}2 \log(\pi \De/d^2) +\frac{d}2 \frac{\Dg}{\De} 
+ \frac{d}2 \wh \f_g   
\int dy \, e^y \,  \Th\left( \frac{y +\Dg/2}{\sqrt{2 \Dg}}  \right) \log  \Th\left( \frac{y-\h +\De/2}{\sqrt{2 \De}}  \right) \ .
\eeq
The value of $\De$ is found by imposing $ds_g/d\De=0$.
Note that it is easy to check that $s_g$ has a {\it minimum} as a function of $\De$. This is due to the analytic continuation to $s\to 0$
that changes the sign of the second derivative of $s_g$~\cite{MPV87}. However, this is not important, because the relevant condition
is that the eigenvalues of the stability matrix, and in particular the replicon mode, should be negative.
It is not very useful to write explicitly the equations for $\De$, that can be easily derived from Eq.~\eqref{eq:sedinfDfzero}. In fact,
numerically it is easier to minimize $s_g$ to obtain the physical value of $\De$.
As in the case $\D^f\neq 0$, when $\h=0$, one finds that $\De = \Dg$ and $s_g$ reduces to the expression of the glass
entropy that is derived in the Monasson calculation~\cite{PZ10}.

For the pressure, from Eq.~\eqref{eq:eqpressure} we obtain when $\D^f\to 0$:
\beq\label{eq:pethermo}
\frac{p_g}d = 
\frac{ \wh\f_g}2 \int dy \, \Th\left( \frac{y +\Dg/2}{\sqrt{2 \Dg}}  \right)
 \frac{ e^{-\frac{(y - \h -\De/2)^2 }{2 \De}}}{\sqrt{2\pi \De}} \left[ \Th\left( \frac{y-\h +\De/2}{\sqrt{2\De}}  \right) \right]^{-1} \ .
\eeq
Note that from this relation it is obvious that when $\wh\f = \wh\f_g$, hence $\h=0$ and $\De = \Dg$, one has $p_g = d \, \wh\f/2$. 
Therefore, as in the case $\D^f\neq 0$, the pressure is continuous when the
glass equation of state merges with the liquid one at $\wh\f = \wh\f_g$.

\subsubsection{Pressure and thermodynamic relations}

For hard spheres the pressure is defined, at the leading order for $d\to\io$, as~\cite{Hansen,PZ10}:
\beq\label{eq:py}
p = - \wh\f \frac{d s}{d\wh\f} = \frac{d \, \wh\f}2 y(\wh\f) \ ,
\eeq
where $y(\wh\f)$ is the contact value of the pair correlation.
For the liquid phase, $y_{\rm liq}(\wh\f)=1$ and $p_{\rm liq} = d \, \wh\f/2$.
Here we want to show that this thermodynamic relation is also satisfied in the glass phase. The proof could be also given in the general
case but we provide it only in this approximated setting with $\D^f=0$ for simplicity. Note that this relation is not satisfied if one performs
an approximate isocomplexity calculation in the Monasson framework~\cite{PZ10}.

To obtain $y(\wh\f)$ we need to
compute the contact value of the correlation function of the glass. 
In $d=\io$, the correlation function coincides with the two-replica effective potential~\cite{PZ10}.
The Gaussian approximation of Sec.~\ref{sec:Gaussian} is exact for $d\to\io$~\cite{KPZ12} and can be used
to derive the effective potential following~\cite{CKPUZ13}. Here we only sketch the derivation without discussing the details 
because the procedure
of~\cite{CKPUZ13} is very easy to generalize.

We consider the Gaussian approximation of Eq.~\eqref{Gaussian} for $m=1$ and $\DE^f=0$.
We have one reference replica interacting with potential $v$ and with cage radius $\DE^g/2$, and $s$ 
constrained replicas with potential $v_\g$ (recall that here $\g$ denotes a generic perturbation that in our case
is the compression)
and $\DE/2$, with $s\to 0$. The effective
interaction of one constrained replica is therefore
\beq
e^{-\b \phi_{\rm eff}(x)} = e^{-\b v_\g(x)} \int dX \g_{\DE}(x-X) \, q_{\DE^g/2}(X) \, q^\g_{\DE/2}(X)^{s-1} \ ,
\eeq
where $q_{\DE^g/2}(x) = \int dy \g_{\DE^g}(x-y) e^{-\b v(x)}$ and $q^\g_{\DE/2}(x)= \int dy \g_{\DE}(x-y) e^{-\b v_\g(x)}$.
Here $v(x)$ is a hard-sphere potential with diameter $D_g=1$ and $v_\g(x)$ is a hard-sphere potential with diameter $D = 1 +\h/d$.
Using bipolar coordinates and taking the limit $d\to\io$ of the integrals~\cite{PZ10,CKPUZ13}, we obtain, for $r=|x|$ and $t = d(r-1)$:
\beq\label{eq:corr}
e^{-\b \phi_{\rm eff}(t)} = \th(t-\h) \int dy \, \frac{ e^{-\frac{(y - t - \De/2)^2 }{2 \De}}}{\sqrt{2\pi \De}}  \Th\left( \frac{y +\Dg/2}{\sqrt{2\Dg}}  \right) 
\left[ \Th\left( \frac{y- \h + \De/2}{\sqrt{2 \De}}  \right) \right]^{s-1} \ . 
\eeq
We have $g_g(r) = e^{-\b \phi_{\rm eff}(r)}$ and therefore $y_g = e^{-\b \phi_{\rm eff}(\h)}$, with $s\to 0$.
The resulting pressure is
\beq\label{eq:pecorre}
p_g = \frac{d \, \wh\f_g}2 y_g = \frac{d \, \wh\f_g}2  \int dy \, \frac{ e^{-\frac{(y - \h - \De/2)^2 }{2 \De}}}{\sqrt{2\pi \De}}  \Th\left( \frac{y + \Dg/2}{\sqrt{2 \Dg}}  \right) 
\left[ \Th\left( \frac{y- \h + \De/2}{\sqrt{2 \De}}  \right) \right]^{-1} \ ,
\eeq
which coincides with the one obtained via the thermodynamic route, Eq.~\eqref{eq:pethermo}. This shows that
Eq.~\eqref{eq:py} is satisfied also in the glass phase, hence {\it the theory is thermodynamically consistent}.

\subsubsection{The jamming limit}

We want to analyze what happens in the jamming limit, where the cage radius $\De \to 0$ and the pressure
$p_g \to \io$.
In this case, because $\D^f=0$, Eqs.~\eqref{eq:appj} reduce to a single condition for $\h$. The first equation,
for $\D^f\to 0$ (and $m=1$), becomes
\beq\label{eq:AAA3}
0 =  \Dg   - \frac{\wh\f_g}2  \int_{-\io}^0 dx \, e^{x + \h } \,
 \Th\left( \frac{x + \h + \Dg/2}{\sqrt{2 \Dg}}  \right) 
x^2 \ .
\eeq
This equation defined the value $\h_j$ which depends on the initial density $\wh\f_g$ and its corresponding
reference radius $ \Dg$. The solution of this equation corresponds to the jamming density of the glass 
prepared at density $\wh\f_g$.

Next we can look at the divergence of the pressure. Starting from Eq.~\eqref{eq:pecorre}, making the same change
of variable $x = y - \h + \frac{\De}{2}$ as before, using 
\beq\label{eq:AAA2}
\frac{e^{- \frac{x^2}{2 \De}   }}{\sqrt{2 \pi \De}} \Th\left( \frac{x}{\sqrt{2 \De}}  \right)^{-1}
\to   \frac{|x|}{ \De} 
\eeq
and keeping the leading order in small $ \De$, we get
\beq\begin{split}
p_g &= 
\frac{d \, \wh\f_g}2  \int dx \, \frac{ e^{-\frac{(x - \De)^2 }{2 \De}}}{\sqrt{2\pi\De}}  
\Th\left( \frac{x + \h - \De/2 + \Dg/2}{\sqrt{2 \Dg}}  \right) 
\Th\left( \frac{x}{\sqrt{2 \De}}  \right)^{-1} \\
&=
\frac{d \, \wh\f_g}{2 \De} \int_{-\io}^0 dx \,|x|
\Th\left( \frac{x + \h + \Dg/2}{\sqrt{2 \Dg}}  \right) e^{x } \ . \\
\end{split}\eeq
Therefore the pressure diverges as $p_g \sim 1/ \De$.

\section{Results}

We present here some additional details on how we obtained the results for the physical observables
discussed above.
For each $\wh\f_g$, we first solve numerically Eq.~\eqref{eq:Dg} to obtain $\Dg$. This can be done very
easily, for example by minimizing the entropy~\eqref{eq:seref}. The integrals converge well and can be easily
handled by many softwares (we used Mathematica).
Then, we solve numerically Eqs.~\eqref{eq:eqD}-\eqref{eq:eqDf} to obtain $\De$ and $\D^f$ as functions
of $\g$ and $\h$. These equations are solved by iteration. However, a little bit of care is needed here
to handle the integrals because of the presence of different scales. We wrote a code in C by discretizing
the integrals using standard techniques. We also made use of the 
``Faddeeva" package\footnote{\url{http://ab-initio.mit.edu/wiki/index.php/Faddeeva_Package}}
to improve the precision of the error functions in the asymptotic regimes.
To test our code, we checked that it reproduces the limiting cases discussed above: for example, the jamming
point $\h_j$ and the corresponding value of $\D^f_j$, and the perturbative behavior at small $\h$ and small
$\g$.

\subsection{Compression}

\begin{figure}[t]
\includegraphics[width=.45\textwidth]{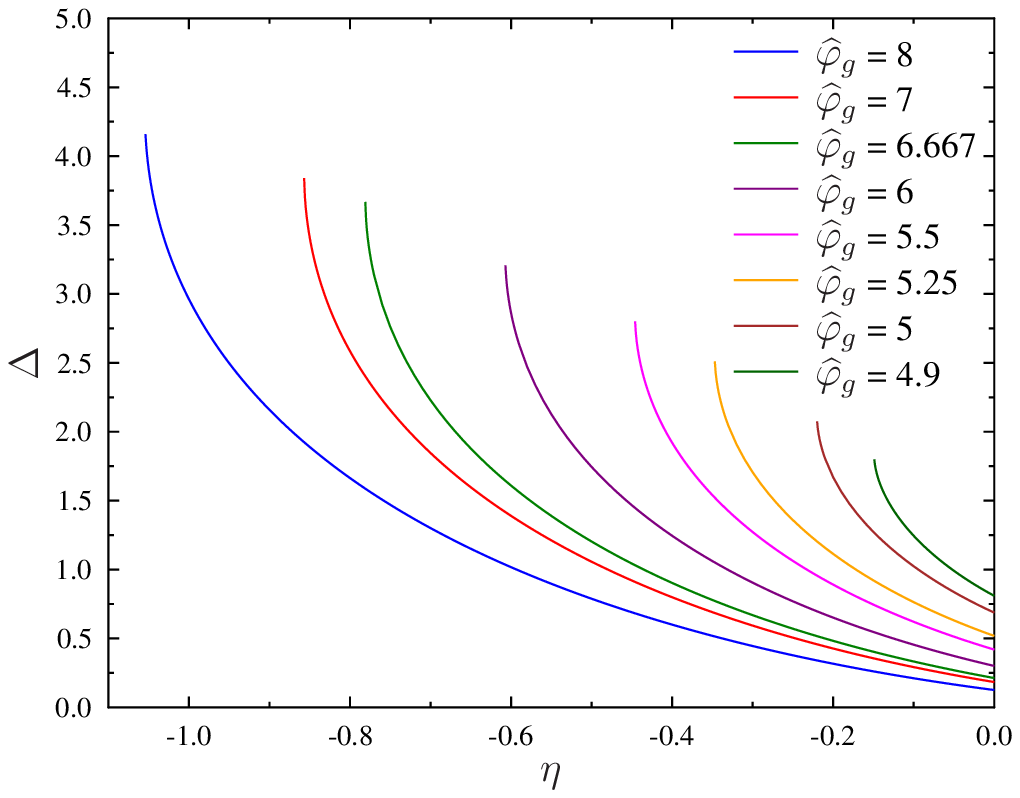}
\includegraphics[width=.45\textwidth]{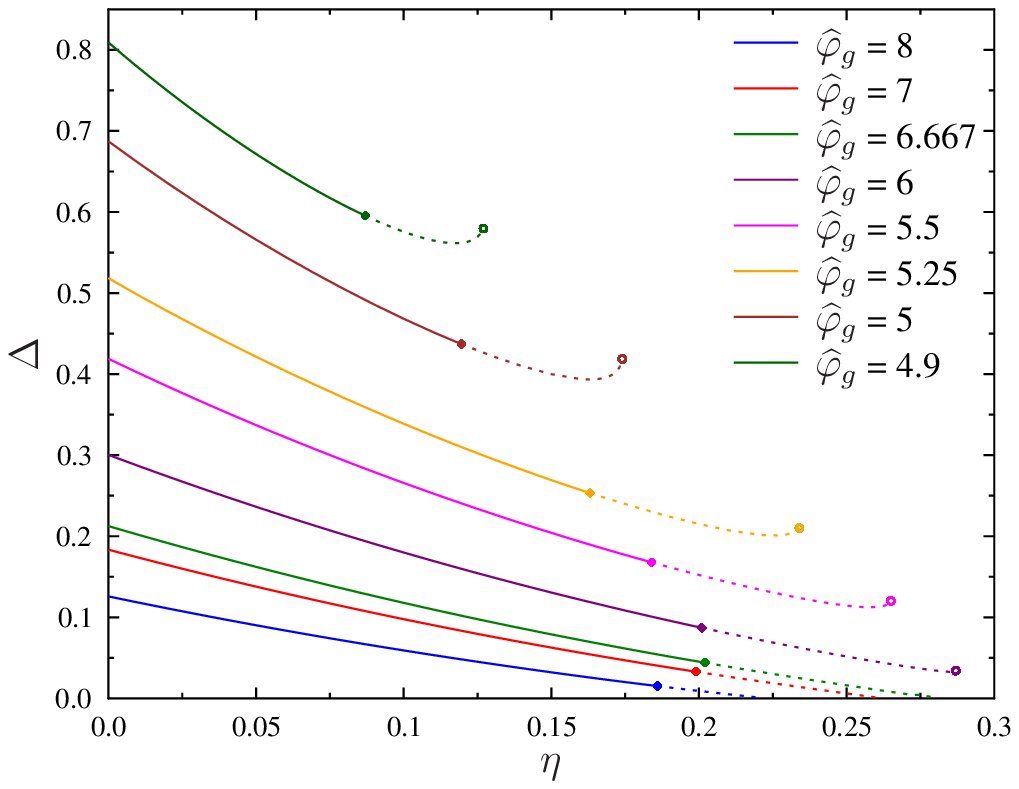}
\includegraphics[width=.45\textwidth]{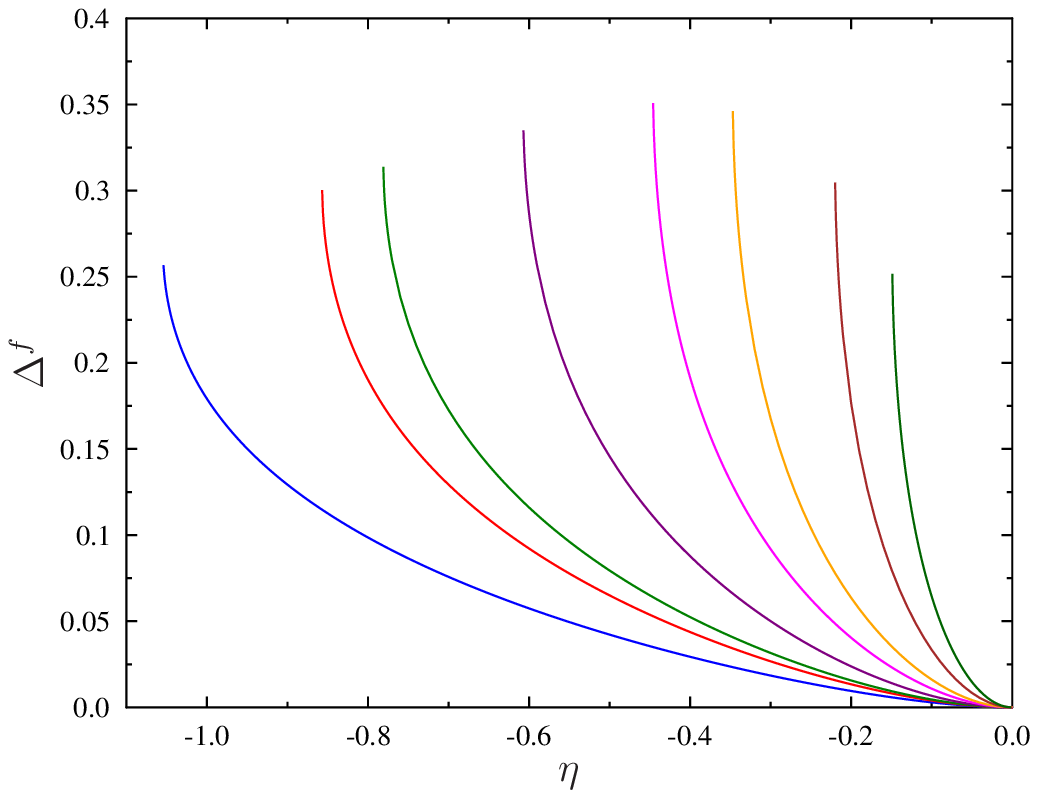}
\includegraphics[width=.45\textwidth]{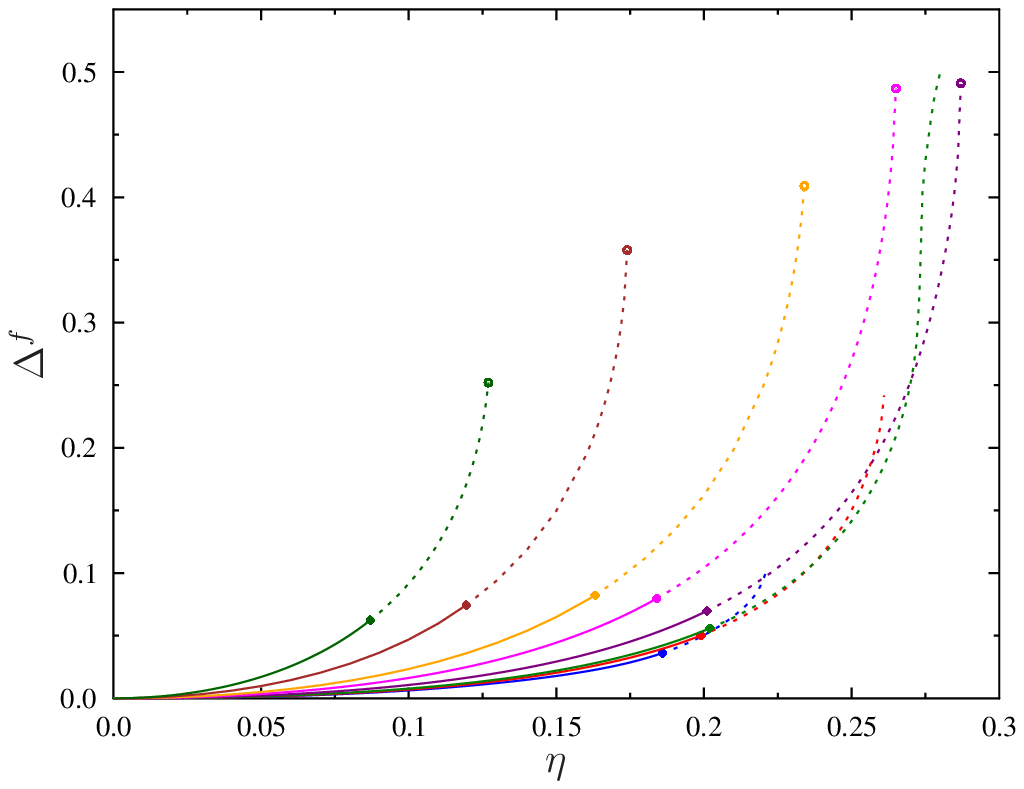}
\caption{
$\De$ ({\it top panels}) and $\D^f$ ({\it bottom panels}), solutions of Eqs.~\eqref{eq:eqD}-\eqref{eq:eqDf},
for different glassy states
followed in decompression ($\h<0$, {\it left panels}) and compression ($\h>0$, {\it right panels}).
We use separate scales to improve the readability of the figures. The dashed lines indicate the unstable
region where the replicon mode is positive (Fig.~\ref{fig:rep_compr}).
}
\label{fig:Delta_compr}
\end{figure}

In Fig.~\eqref{fig:Delta_compr} we report the evolution of $\De$ and $\D^f$ under compression ($\h>0$) or
decompression ($\h<0$) at $\g=0$. The corresponding pressure is reported in Fig.1 of the main text.
In decompression, we find that $\D^f$ increases quadratically from zero, while $\De$ also increases. 
At low enough $\h$, a spinodal point is met, where the solution disappears. This is signaled by a square-root singularity
in both $\De$ and $\D^f$, as usual for spinodal points. At that point the glass ceases to exist and melts into the liquid phase.

In compression, again $\D^f$ increases quadratically while $\De$ decreases. At high enough $\wh\f_g$
(see e.g. $\wh\f_g=8$ in Fig.~\ref{fig:Delta_compr}), $\De$ vanishes linearly at $\h_j$, while $\D^f$ is finite at $\h_j$,
as predicted by the asymptotic analysis of Sec.~\ref{sec:jamming}. The values of $\h_j$ and $\D^f_j$ coincide with the ones
obtained using the analysis of Sec.~\ref{sec:jamming}.
At low density (see e.g. $\wh\f_g=5$ in Fig.~\ref{fig:Delta_compr}), however, before the jamming point, an unphysical spinodal point is reached (signaled again by a square root singularity),
marked by a symbol in Fig.~\ref{fig:Delta_compr}.
This unphysical spinodal point has also been found in spin glasses~\cite{KZ10b}.

\begin{figure}[h]
\includegraphics[width=.45\textwidth]{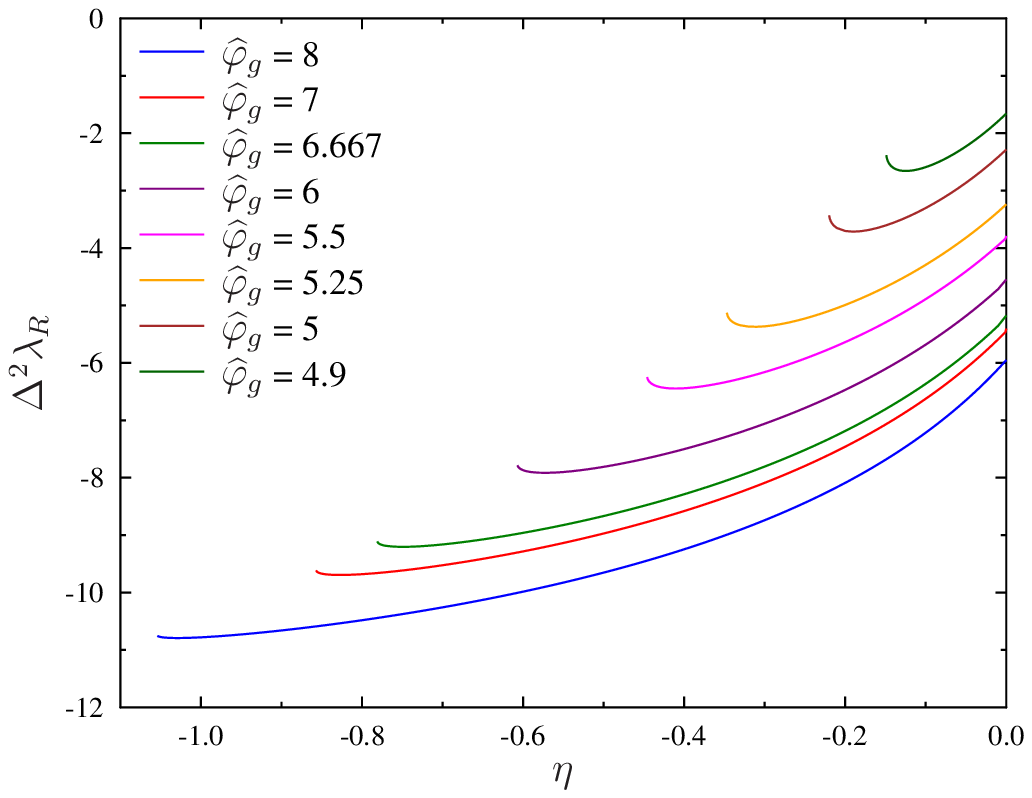}
\includegraphics[width=.45\textwidth]{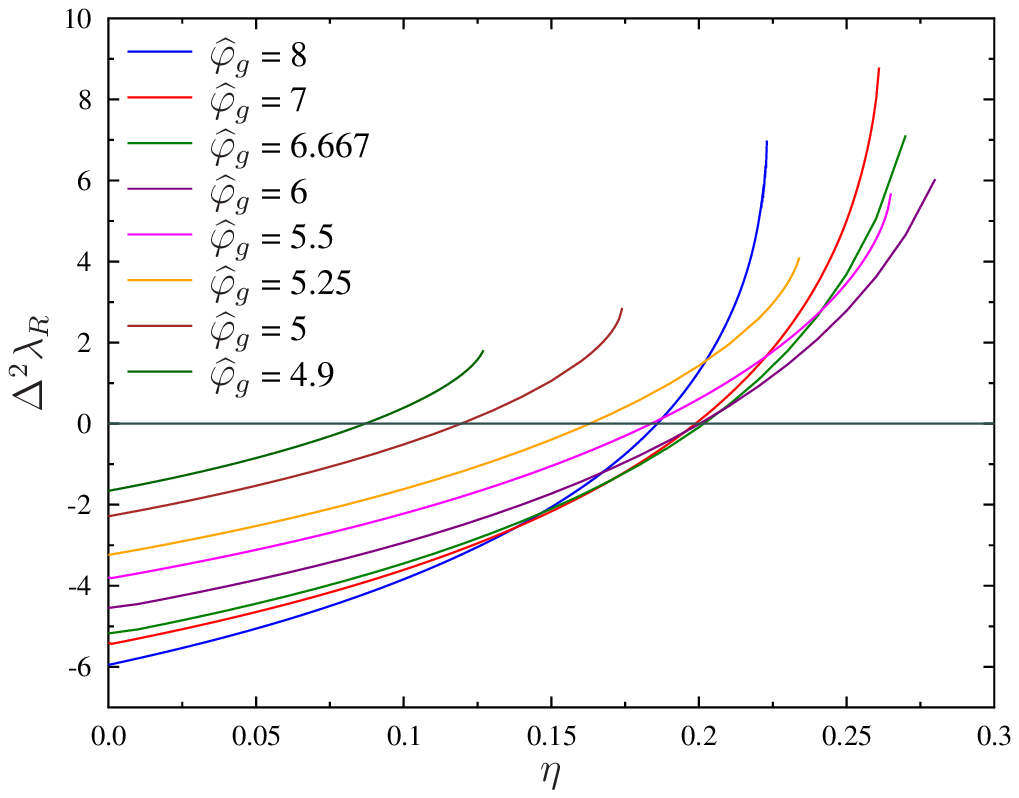}
\caption{
The replicon mode given by Eq.~\eqref{eq:lR}, for the same glasses as in Fig.~\ref{fig:Delta_compr} and Fig.1 of the main text.
In decompression the replicon is always negative, therefore the RS solution is stable; in compression, the replicon vanishes at the Gardner
transition signaling an instability of the RS solution.
}
\label{fig:rep_compr}
\end{figure}

In both cases, before either jamming or the unphysical spinodal point is reached, the replicon mode becomes positive (Fig.~\ref{fig:rep_compr}),
signaling that the glass state undergoes a Gardner transition~\cite{Ga85,CKPUZ13} and beyond that point the RS solution we used is not correct.
Based on the analogy with spin glasses and in particular on the results of~\cite{KZ10b}, and on the discussion of Sec.~\ref{sec:replicongen},
we expect that at the Gardner transition replica symmetry is broken towards a fullRSB solution. A 1RSB structure in the $s$-block should already
give an excellent approximation of the equation of state of the glass and eliminate the unphysical spinodal point. We leave this computation for
future work. The unstable part of the curves in Fig.~\ref{fig:Delta_compr} is reported with dashed lines.

The result for shear modulus, reported in Fig.3 of the main text, 
is easily deduced from the results for $\De$ reported in Fig.~\ref{fig:Delta_compr} using Eq.~\eqref{eq:mu}.
We can also compute the dilatancy from Eq.~\eqref{eq:R}: the result is reported in Fig.~\ref{fig:dilatancy}.
Note that $R/\rho=(1/2)\wh\f\partial \mu/\partial \wh\f$ as it can be deduced by combining Eqs.~\eqref{eq:R} and \eqref{eq:mu}.
From Eq.~\eqref{eq:R} it is also clear that $R$ diverges at the spinodal point where $\De$ has a square-root singularity (hence
infinite derivative) and at the jamming point where $\De\to 0$.

\begin{figure}[b]
\includegraphics[width=.45\textwidth]{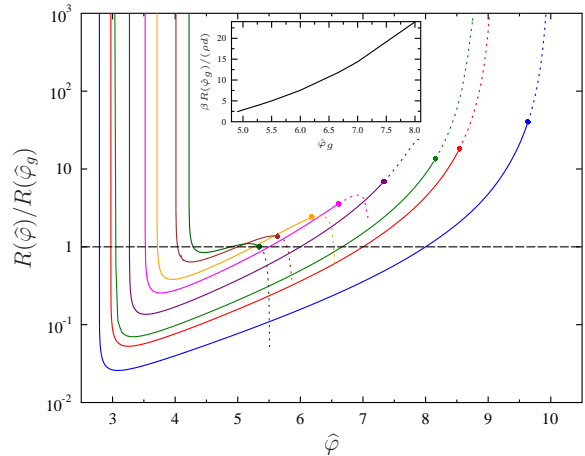}
\caption{
Dilatancy $R$ as a function of density for different glasses. Colors and styles as in Fig.~3 of the main text where the
shear modulus $\mu$ is reported. Recall that $R/\rho=(1/2)\wh\f\partial \mu/\partial \wh\f$.
In the inset, the evolutions of $R(\wh\f_g)$ with $\wh\f_g$ is reported. 
Note that the dilatancy diverges both at jamming and at
the low density spinodal point where the glass melts.
}
\label{fig:dilatancy}
\end{figure}

\subsection{Shear}

\begin{figure}[b]
\includegraphics[width=.45\textwidth]{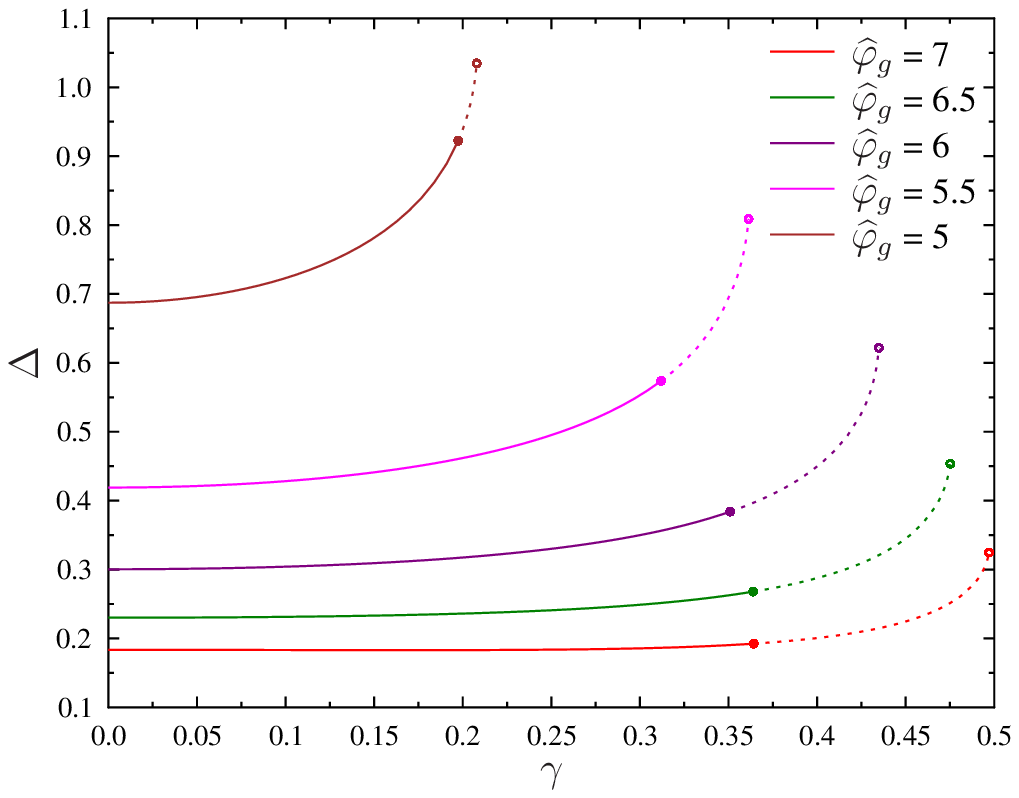}
\includegraphics[width=.45\textwidth]{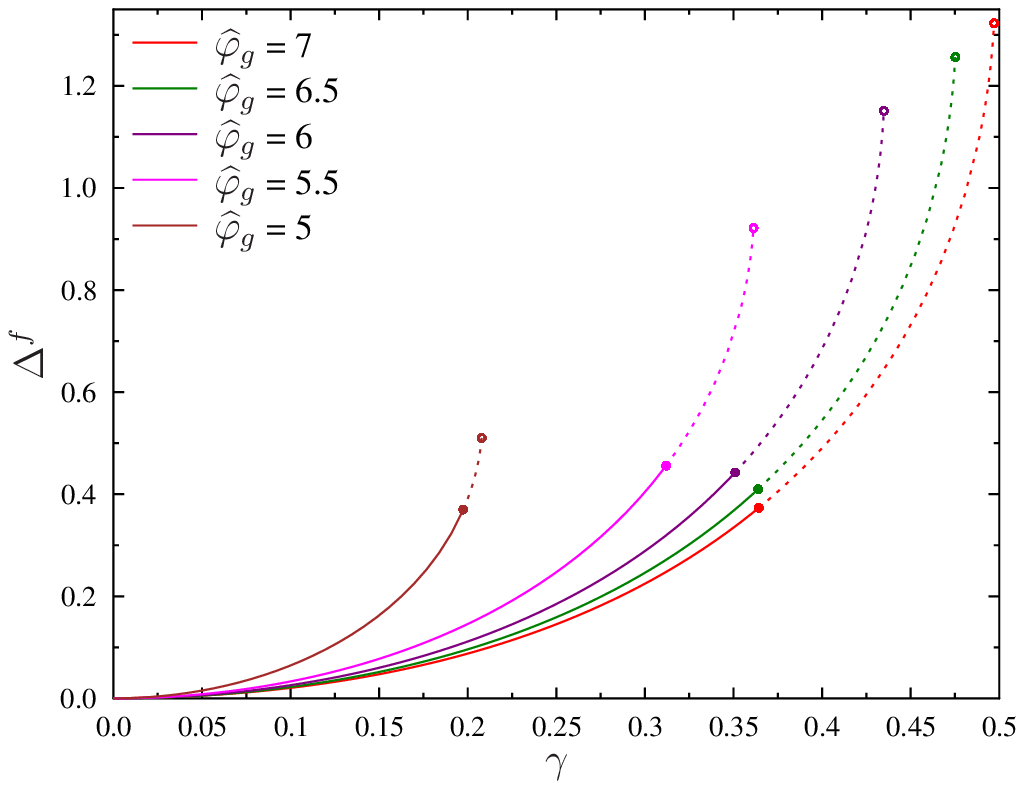}
\caption{
Values of $\D$ ({\it left panel}) and $\D^f$ ({\it right panel}) as functions of shear strain $\g$ in absence of compression ($\h=0$).
 The dashed lines indicate the unstable
region where the replicon mode is positive (Fig.~\ref{fig:rep_shear}).
}
\label{fig:Delta_shear}
\end{figure}

We next analyze the behavior under a shear strain $\g$ in absence of compression, hence for $\h=0$.
The results for $\De$ and $\D^f$ are reported in Fig.~\ref{fig:Delta_shear}. We observe that upon increasing
$\g$ both $\De$ and $\D^f$ increase, until a spinodal point is reached, at which they both display a square root singularity.
Correspondingly, both the shear stress $\s$ and the glass pressure $p_g$ (main text, Fig.2) display a square root singularity.

However, before the spinodal is met, the replicon mode becomes positive (Fig.~\ref{fig:rep_shear}) and the system undergoes
a Gardner transition.
The fact that the a Gardner transition is met when the system is subject to a shear strain 
might be surprising at first sight, because one could think that straining a well defined glass basin amounts to deform the basins
but should not induce its breaking into sub-basins. 
However, note first that on general grounds, 
the free energy landscape can change once perturbations are added.
Moreover, we find (see Fig.2 in the main text) that the pressure of the glassy state increases when the shear strain is increased.
This means that under shear strain the particles in the glass basins 
become more constrained
and because of this some parts of the basin can become forbidden, triggering the Gardner transition as it happens during a compression
in absence of shear strain.

\begin{figure}[h]
\includegraphics[width=.45\textwidth]{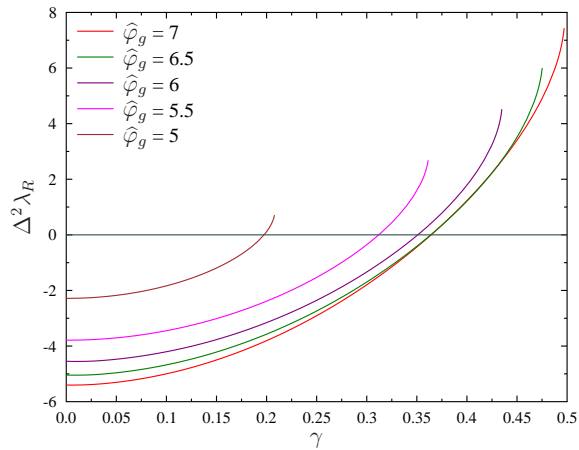}
\caption{
The replicon mode given by Eq.~\eqref{eq:lR}, for the same glasses as in Fig.~\ref{fig:Delta_shear} and Fig.2 of the main text.
The replicon vanishes at the Gardner
transition signaling an instability of the RS solution.
}
\label{fig:rep_shear}
\end{figure}

\subsection{Quality of the approximation $\D^f=0$}

To conclude, we comment on the quality of the approximation $\D^f$ discussed in Sec.~\ref{sec:appr0} at $\g=0$.
We see from the results of Fig.~\ref{fig:Delta_compr} that this approximation is reasonable, except close to the spinodals
where $\D^f$ grows rapidly. Correspondingly, we find that the approximation $\D^f=0$ misses the spinodals. In decompression,
the states can be followed to arbitrarily low density, which is clearly unphysical. In compression, the spinodal is missed, but the Gardner
transition is predicted with quite good accuracy, hence the approximation does a fair job in the physical region.

\section{Comparison with previous work}

Let us conclude by a comparison with previous work~\cite{PZ10,KPZ12,KPUZ13,CKPUZ13}, in which
the Monasson formalism was used~\cite{Mo95}. Let us illustrate shortly the outcome of the Monasson computation.
We restrict to hard spheres for simplicity. At a given packing fraction $\f$ in the glass phase, the phase space of the system is decomposed
in a certain number of glassy states, each glass state being a {\it cluster} of configurations (for an analysis of this clustering phenomenon in a
similar context see~\cite{KMRSZ07}). Each cluster contains a certain number of configurations, which defines its internal entropy $s$.
The {\it complexity} $\Si(s) = N^{-1} \log \NN(s)$ is the logarithm of the number of clusters that have entropy $s$, at fixed density $\f$.
This is what is computed by the Monasson formalism (through the introduction of a parameter $m$ conjugated to the internal entropy of the cluster).
The crucial point, however, is the following. Consider all the clusters with total entropy $s$. Among them, there are clusters with different properties
(e.g. pressure, shear modulus, etc.). However, since the total number of clusters is exponential in $N$, as usual in statistical mechanics,
there is a subset of them (we call them {\it typical}) that have {\it typical} properties (same pressure, same shear modulus, etc.). The Monasson
formalism allows one to compute the properties of these typical states. Hence, the resulting pressure-density phase 
diagram reported in~\cite{PZ10,CKPUZ13} refers to the properties of the typical states.

On the other hand, it is well known in spin glasses~\cite{MPV87} that if one takes some states that are typical at a given state point
(i.e. at a given density and internal entropy, or equivalently, at a given density and pressure) and follows their evolution at a different
state point (e.g. at a different density), they become {\it atypical}. This means that they have in general {\it different} values of thermodynamic
observables with respect to the typical states at the new state point. 
The Franz-Parisi formalism~\cite{FP95,BFP97} (or ``state following'' formalism~\cite{KZ10b,KZ13} allows one to select a typical glass
state in some state point, and follow its evolution to a different state point. In this paper we have focused on states that are typical
at a given density $\f_g$ and at the value of pressure corresponding to the equilibrium liquid pressure at density $\f_g$ (that for
infinite-dimensional hard spheres is just $p/d = \wh\f_g/2$), and we followed their evolution in compression and in shear.
In the Monasson formalism, these states are selected by choosing $m=1$. Choosing different values of $m$ (here we wrote all the equations for general
$m$ even if in the end we only considered $m=1$) allows one to select
different states, and then we can follow them to a different state point using the formalism we developed above.

The most interesting and striking difference with respect to the typical computation~\cite{KPUZ13,CKPUZ13} (see also~\cite{Ri13} for spin glasses) 
concerns the behavior of the Gardner 
transition line (main text, Fig.1) around the dynamical transition.
In the present work, we showed that
for states prepared at $\f_g = \f_{\rm d}$,
the Gardner transition is met immediately after an infinitesimal compression (see also~\cite{BFP97,FPR14}).
In other words, the Gardner transition line ends 
at the dynamical transition (see main text, Fig.1).
This can be understood heuristically by observing that glasses with $\f_g \sim \f_{\rm d}$ correspond to fast compression procedures, 
while glassy states with $\f_g \gg \f_{\rm d}$ correspond to very slow compression. 
It is therefore reasonable that the former are more unstable than the latter. 
However, in \cite{Ri13,KPUZ13} it was found that the fullRSB phase appears only above a given packing fraction 
$\varphi^*> \varphi_d$ implying that all states around the dynamical point are stable and no fullRSB is present around $\f_{\rm d}$.
The reason behind this difference is that
the equilibrium states, once followed in different state points, 
become immediately atypical. The states prepared at $\f_{\rm d}$ undergo a Gardner transition immediately under compression, 
but they become atypical and are not detected by the Monasson
computation~\cite{KPUZ13}, which find instead other glassy states that {\it appear} away from equilibrium and are stable until $\f < \f^*$.
This is a signal that the evolution of the free energy landscape under external perturbations is very chaotic in the region around $\f_{\rm d}$.

\clearpage

\end{widetext}

\bibliographystyle{mioaps}
\bibliography{HS}

\end{document}